\documentclass[conference]{IEEEtran}
\IEEEoverridecommandlockouts
\usepackage{cite}
\usepackage{amsmath,amssymb,amsfonts}
\usepackage{graphicx}
\usepackage{textcomp}
\usepackage{xcolor}
\usepackage{comment}
\def\BibTeX{{\rm B\kern-.05em{\sc i\kern-.025em b}\kern-.08em
    T\kern-.1667em\lower.7ex\hbox{E}\kern-.125emX}}
\usepackage{algorithm}
\usepackage{algpseudocode}
\usepackage{booktabs}
\usepackage{hyperref}

\begin{document}

\newcommand{\sys}{\textsc{Cutpurse}}

\title{\sys: Decomposing Explanation-Guided Model Extraction on Tabular Data
}

\author{\IEEEauthorblockN{Seyma Selcan Magara\textsuperscript{*}, Hanru Zhang\textsuperscript{*}, Fatme El Zahraa Hamad, Nico Pfeifer, Mete Akg\"un\textsuperscript{*}}
\IEEEauthorblockA{\textit{University of T\"ubingen}, T\"ubingen, Germany \\
\textsuperscript{*}\textit{Medical Data Privacy and Privacy-Preserving Machine Learning Group} \\
\{seyma-selcan.magara, hanru.zhang, nico.pfeifer, mete.akguen\}@uni-tuebingen.de \\
fatme-el-zahraa.hamad@student.uni-tuebingen.de}
}

\maketitle

\begin{abstract}
Explanation APIs are often treated as increasing model-extraction risk, but an end-to-end attack comparison cannot show whether an explanation reveals model-specific information or merely supplies a convenient next-query coordinate. We study this distinction for explanation-guided extraction of tabular classifiers. We separate an explanation into an \emph{attribution} channel, which ranks influential features, and a \emph{coordinate} channel, which supplies candidate feature values for perturbing those features. We then test whether the useful bin edges are privileged to the service or reproducible from the adversary's own data.

Across six datasets and five target families, using attribution rankings to choose which features to perturb provides no systematic extraction advantage. \textsc{Lime}'s discretization bin edges can improve fidelity by up to $41$ percentage points on continuous features, but those edges are quantiles of the explainer's background data rather than properties of the target model. In controlled comparisons, an attacker-side grid reproduces the useful bin-edge gains without querying an explanation API. We use these findings to derive \sys, which combines a self-computed grid with diverse query generation and boundary search. Against a service that withholds bin edges, \sys\ improves fidelity by $16.3$ percentage points on continuous-feature targets. It also exceeds the undefended published attack by $7.0$ percentage points. Thus, withholding default \textsc{Lime} bin edges alone does not prevent effective extraction under the in-domain auxiliary-input, probability-output, and same-family surrogate assumptions studied here. Our results concern default \textsc{Lime} and \textsc{Shap} configurations for tabular classifiers. They do not establish the same conclusion for other explainers, target classes, or data-free attackers.
\end{abstract}


\section{Introduction}
Machine learning services increasingly return a post-hoc explanation alongside each prediction. This practice improves transparency, but it also raises a security question: does an explanation endpoint give a model-extraction adversary a capability that prediction access alone cannot provide? Model extraction builds a surrogate that agrees with a black-box target on unseen inputs~\cite{tramer2016stealing,jagielski2020high}. High-fidelity extraction compromises model confidentiality and can support offline transfer-based attacks~\cite{tramer2016stealing,papernot2017practical,yue2021blackbox}. It can also amortize the target-query cost of downstream membership inference~\cite{oksuz2026lomime}.

We take Autolycus~\cite{oksuz2024autolycus}, an explanation-guided extraction attack against interpretable tabular classifiers, as our starting point. It couples the feature information returned by \textsc{Lime} or \textsc{Shap} with a traversal that generates new queries. Its evaluation shows that this complete attack can extract models effectively under a limited query budget. It does not, however, isolate how much the explanation outputs (feature rankings and LIME bin edges) and the traversal each contribute to that performance. \textsc{Lime} provides both a feature ranking and bin edges, while \textsc{Shap} provides feature attributions without bin edges. Even if an explanation output helps, it remains unclear whether it reveals information specific to the service or information the attacker can approximate from its own data. These distinctions determine whether the explanation API provides a model-specific capability or merely a convenient query-placement aid.

We decompose the information Autolycus consumes into two channels. The attribution channel ranks influential features. The bin-edge channel supplies candidate feature values for perturbing those features. This decomposition lets us test each channel while holding the rest of the attack fixed. It also gives a controlled test for a service-specific disclosure. We compare the service's bin edges with equivalent edges that the adversary estimates from its own auxiliary data.

The resulting diagnosis also motivates \sys. Autolycus explores outward from a small seed set, so it can miss regions that no seed reaches. It also does not actively search between opposite-class samples for a decision crossing. We address these limitations with diverse query generation and boundary search. For bin-edge placement, \sys\ fits \textsc{Lime}'s default discretization grid to its own auxiliary inputs. It combines this auxiliary grid with the two traversal components and never calls an explanation API. Section~\ref{sec:method} derives these three design choices.

\paragraph{Contributions}
\begin{itemize}
  \item \textbf{A channel decomposition of explanation-guided extraction.} We separately evaluate feature selection by attribution ranking and \textsc{Lime} bin edges in the base Autolycus traversal. Across six datasets and five target families, ranking-based selection provides no systematic advantage, whereas bin edges can improve fidelity by up to $41$ percentage points on continuous-feature targets (\S\ref{subsec:e2}--\ref{subsec:e3}).
    \item \textbf{A controlled test of service-specific bin-edge information} For LIME’s default quartile discretizer, bin edges are quantiles of the explainer’s background data, not properties of the target model. Under the auxiliary-input conditions studied, attacker-fitted grids reproduce similar bin-edge gains. Grids fitted on small initial seed pools retain most of these gains. A sensitivity check with LIME’s label-dependent entropy discretizer gives the same qualitative result.
  \item \textbf{A constructive explanation-free attack.} \sys\ combines an attacker-computed grid with established diversity and boundary-search techniques and makes no explanation calls. On continuous-feature targets, it improves fidelity by $16.3$ percentage points over the defended Autolycus configuration and by $7.0$ percentage points over Autolycus with bin edges available (\S\ref{subsec:ladder}). It demonstrates that an adaptive attacker can recover useful query-placement capability independently of the explanation API under the studied assumptions.

\end{itemize}

\paragraph{Scope} We study one family of explanation-guided attacks on interpretable tabular classifiers. Our results do not imply that explanations are harmless in general. These conclusions may not extend to other explanation methods or configurations, high-dimensional or neural targets, or data-free attacks.


\section{Background and Related Work}
\label{sec:background}

\subsection{Model Extraction and Post-hoc Explanations}
\label{subsec:prelim}

Model extraction attacks reconstruct a functional copy $\hat{f}$ of a black-box target $f$ from query access alone, with success measured by prediction agreement between $\hat{f}$ and $f$ on held-out data (formalised in \S\ref{sec:problem}). When the target is deployed as an explainable service, each query returns not only a label but a post-hoc explanation, which may reveal additional structure about $f$. Two dominant explanation families expose fundamentally different information:

\begin{itemize}
  \item \textbf{Shapley-based feature attribution (SHAP)} SHAP~\cite{lundberg2017shap} assigns each feature a signed magnitude $\phi_i(x)$, its Shapley marginal contribution to the prediction at $x$. This is a statement of how much each feature matters locally. It carries no explicit information about where along a
  feature axis the decision changes.
  \item \textbf{Local surrogate explanation (LIME)} LIME fits an interpretable model to a discretized neighbourhood of $x$. Continuous features are binned, and the returned coefficients are attached to $(\text{feature}, \text{bin})$ pairs. LIME therefore reveals both an influential feature and the bin edge of the local discretization bin.
\end{itemize}

This distinction between feature rankings alone and feature rankings accompanied by
local bin edges is central to our analysis. We test whether the base Autolycus traversal benefits from using an attribution ranking for feature selection and whether it benefits from LIME bin edges (\S\ref{sec:problem}, \S\ref{sec:analysis}).

\subsection{Related Work}
\label{subsec:related}

\paragraph{Model Extraction Attacks}
Model extraction attacks aim to reconstruct a functional copy of a black-box target model through strategic querying~\cite{tramer2016stealing}.
Early work by Lowd and Meek framed adversarial classifier reverse engineering from query access~\cite{lowd2005adversarial}.
Tram\`er~et~al.~\cite{tramer2016stealing} demonstrated that
simple equation-solving over prediction outputs suffices for linear models.
Subsequent methods scaled to neural networks: Knockoff Nets~\cite{orekondy2019knockoff}
trains a surrogate on outputs obtained by querying with a knock-off dataset,
while Jagielski~et~al.~\cite{jagielski2020high} show that high-fidelity extraction
of deep networks is possible given sufficient queries.
Other work has examined extraction from complementary perspectives, framing it as knowledge distillation~\cite{papernot2017practical,correia2018copycat} or active learning~\cite{chandrasekaran2020active}, where queries are selected to efficiently transfer or recover the target model's decision boundary.
This line of work has also motivated the defence methods, for example, Juuti~et~al.~\cite{juuti2019prada} propose PRADA, which detects anomalous query distributions
associated with DNN model-stealing attempts.
More recently, Marich~\cite{karmakar2023marich} formalises the problem as distributional equivalence
and proposes an active sampling strategy that maximises information gain per query.
In contrast, our method requires only a small per-class seed set rather than a large
public transfer set, and synthesises informative queries from that seed.

\paragraph{Explanation-Guided Extraction}
The observation that post-hoc explanations leak structural information about the
target model motivates a distinct attack family.
Milli~et~al.~\cite{milli2019model} first demonstrated that gradient-based explanations
suffice to reconstruct linear models exactly.
Miura~et~al.~\cite{miura2024megex} (MEGEX) carry gradient explanations into the data-free setting: treating the returned gradient as a derivative signal, the adversary trains a generative
model that synthesises its own queries, so no real input data is needed. Both exploit a property our
\textsc{Shap} and \textsc{Lime} setting lacks, namely that a gradient specifies a direction in
input space rather than only a per-feature magnitude.
Autolycus~\cite{oksuz2024autolycus} extended this to classical ML classifiers
(decision trees, logistic regression, naive Bayes, and $k$-NN) by using LIME and SHAP
explanations as traversal guides: the attacker perturbs seed samples along the
directions of highest feature attribution, progressively labelling a synthetic dataset
to train a surrogate. Yan~et~al.~\cite{yan2023explanationleaks} (XaMEA) similarly
exploit the spatial information contained in explanations in a model-agnostic manner.
Our question is narrower than whether explanations help extraction in general. For the default LIME and SHAP configurations used by Autolycus, we separately test the attribution ranking and LIME bin edges. We also test whether the bin edges supplied by the service outperform a grid fitted by the adversary.

\paragraph{Counterfactual-Based Extraction}
A growing body of work exploits counterfactual explanations (CFs) rather than
attribution scores.
A\"ivodji~et~al.~\cite{aivodji2020model} first showed that CFs leak enough
boundary information to build high-fidelity surrogates, and DualCF~\cite{wang2022dualcf}
improved query efficiency by training on CF/counter-CF pairs that
straddle the decision boundary.
Khouna~et~al.~\cite{khouna2025tra} propose the Tree Reconstruction Attack (TRA),
which uses a restricted CF oracle to perform axis-aligned binary search,
achieving exact functional reconstruction of decision trees and random forests.
Although theoretically elegant, TRA requires an oracle that returns the
nearest counterfactual confined to a specific hyper-rectangle, an assumption rarely satisfied by deployed XAI APIs.
Concurrently, Otto et al.~\cite{otto2026linear} derive query complexity bounds for extracting
linear classifiers under factual and counterfactual query regimes.
Our setting does not assume a counterfactual oracle. \sys\ first constructs cross-class pairs from its query pool, then uses prediction queries to bisect between them. LIME bin edges are used only as candidate values when constructing those queries. They are not treated as counterfactuals or as boundary locations.

\paragraph{Privacy Implications of Explanations}
Recent work highlights the security cost of XAI transparency.
Shokri~et~al.~\cite{shokri2021privacy} show that feature-based explanations can increase
membership-inference risk by revealing information about training examples.
Ezzeddine et al.~\cite{ezzeddine2026membership} show that exposing counterfactual explanations through query-based
APIs makes shadow-based membership inference attacks (MIAs) more effective in MLaaS settings, and
characterise a three-way trade-off between privacy leakage, predictive performance and explanation
quality.
Quan et al.~\cite{quan2022amplification} make the complementary point that post-hoc feature-importance explanations can act as a side channel, lowering the query cost of evasion, membership inference, and extraction. Such risks motivate privacy-preserving explanation mechanisms, surveyed by Nguyen et al.~\cite{nguyen2024survey}.
We focus on model confidentiality, measuring surrogate fidelity rather than a downstream inference attack. Our controlled grid comparison asks whether withholding LIME bin edges provides protection beyond what an adversary can reproduce from auxiliary inputs.
\section{Problem Formulation}
\label{sec:problem}

\paragraph{Setting}
Let $f : \mathcal{X} \to \mathcal{Y}$ be a black-box target classifier, with $\mathcal{X} \subseteq \mathbb{R}^d$ and $\mathcal{Y} = \{0, \ldots, C{-}1\}$I. The attacker does not receive the target parameters or training data. Following Autolycus, our evaluation trains the surrogate in the target's model family, adopting its assumption that the model type can be identified during retraining~\cite{papernot2016transferability,oksuz2024autolycus}. Each prediction query $x \in \mathcal{X}$ returns a label $f(x)$ and a class-probability vector $\hat{p}(x) \in \Delta^{C}$, and counts against a total budget $Q$.

Our controlled Autolycus experiments additionally expose a post-hoc explanation $e(x)$ from $\mathcal{E} \in \{\textsc{Shap},\textsc{Lime}\}$. \sys\ is evaluated with the
same prediction access but makes no explanation calls. In both cases, the attack starts from the class-balanced seed protocol used by Autolycus: $n$ auxiliary inputs per class ($n{=}1$ for
\textsc{Lime} and $n{=}5$ for \textsc{Shap}). The target labels, the seed queries and those labels are used to train the surrogate.

\paragraph{Security property}
The property under attack is model confidentiality. The attacker seeks a surrogate that agrees with $f$ on held-out inputs. A higher-fidelity surrogate can support downstream inference attacks~\cite{oksuz2026lomime,nguyen2024survey}, but evaluating those attacks is beyond the scope of this work.

\paragraph{Explanation oracle}
Starting from seed queries, baseline work expands its query pool by perturbing samples it
has already queried. Autolycus uses the explanation's feature ranking to choose which input features to perturb. When using LIME, the returned bin edges also provide candidate values for those perturbations. The two explainers therefore expose structurally different information about $f$ near $x$:
\begin{itemize}
  \item \textbf{\textsc{Shap}}: $e(x) = \boldsymbol{\phi}(x) \in \mathbb{R}^{d}$, a vector of
 signed Shapley attributions in which $|\phi_i(x)|$ measures the local influence magnitude of feature $i$~\cite{lundberg2017shap}. It reports how much each feature matters, but does not provide a candidate feature value for the next query. 
  \item \textbf{\textsc{Lime}}: $e(x) = \{(i,\, [l_i, u_i],\, w_i)\}_{i \in F}$ over a selected feature subset $F$, where feature $i$ has local weight $w_i$ and lies in the discretization bin $[l_i, u_i]$. Beyond the feature weight, \textsc{Lime} therefore supplies bin edges that can be used as candidate feature values for placing the next query.
\end{itemize}

\paragraph{Objective}
The attacker trains a surrogate $\hat{f} : \mathcal{X} \to \mathcal{Y}$ that is functionally equivalent to $f$, measured by prediction agreement:
\begin{equation}
    \mathrm{Sim}(\hat{f}, f) \;=\; \mathbb{E}_{x \sim P_{\mathcal{X}}}
    \bigl[\mathbf{1}(\hat{f}(x) = f(x))\bigr].
    \label{eq:sim}
\end{equation}
Here $P_{\mathcal{X}}$ is the (unknown) data distribution over $\mathcal{X}$, estimated in practice by a held-out test set disjoint from the attacker's seed set $\mathcal{D}_{0}$ and its queries.
The attack succeeds if $\mathrm{Sim}(\hat{f}, f)$ is high while the number of API calls stays within $Q$.

\paragraph{What does an explanation leak?}
For each queried sample, the base Autolycus traversal uses the explanation in two ways when it constructs its next perturbations:
\begin{enumerate}
  \item \textbf{Attribution ranking:} a ranking of locally influential features by $|\phi_i|$ or $|w_i|$, available from both \textsc{Shap} and \textsc{Lime}.
  \item \textbf{LIME bin edges:} the edges $\{l_i,u_i\}$ of the local discretization bin. These give candidate values for changing a continuous feature. SHAP magnitudes do not.
\end{enumerate}

We test these inputs and the source of the bin edges with three questions:
\begin{description}
  \item[\textbf{RQ1.}] Does the attribution ranking improve the base traversal over random feature selection?
  \item[\textbf{RQ2.}] Do LIME bin edges improve that traversal over fixed steps from the current value?
  \item[\textbf{RQ3.}] Can an attacker fit an auxiliary grid that substitutes for the target's LIME grid?
\end{description}
RQ1 and RQ2 isolate what the base traversal receives from an explanation. RQ3 holds that traversal fixed and varies only the grid source, so it tests whether the service discloses information the
attacker cannot reproduce.

\paragraph{Auxiliary Data Assumptions}
Autolycus allows the attacker's auxiliary dataset to range from a small set of initial examples to a comprehensive dataset~\cite{oksuz2024autolycus}. In its evaluation, each benchmark dataset is split into target-training ($75\%$), evaluation ($15\%$), and auxiliary ($10\%$) partitions. It then draws $n$ inputs per class from the auxiliary partition to initialize the traversal. We follow this seed protocol: the target is trained only on the target-training partition, and the attacker queries the $n$ initial seeds per class.

Beyond initialization, Autolycus uses auxiliary feature scales to set the perturbation steps in its SHAP traversal. RQ3 asks a different question: can the attacker fit its own LIME grid instead of receiving bin edges from the service? We first fit the attacker grid on the full $10\%$ auxiliary partition to make the main comparison baseline-compatible. This condition fits the attacker-side grid on the full auxiliary partition, although the traversal queries only its initial seeds and generated inputs. 

Because access to a full auxiliary partition is a generous attacker condition,
we use it first to provide a baseline-compatible comparison with Autolycus.
RQ3 then evaluates stricter conditions. The grid-data sensitivity analysis fits the grid using only the same $n$ initial seeds per class usd to start the attack. The class-support sensitivity restricts both the seeds and grid-fitting inputs to a subset of classes, while retaining all classes in the target and evaluation data.
For entropy discretization, the attacker uses the target predictions returned for the initial seed queries as discretizer labels.
\section{Three Weaknesses of Explanation-Guided Traversal}
\label{sec:analysis}
Autolycus expands a small per-class auxiliary seed set by repeatedly explaining the current sample, perturbing its top-ranked features, and querying the results. We identify three limitations of this design. The first two concern the traversal and motivate the two components \sys\ adds (\S\ref{sec:method}). The third concerns the information supplied by the explanation and motivates channel analysis that follows. \S\ref{sec:experiments} evaluates all three.

\paragraph{W1: coverage is a property of the seed set, not the budget}
Covering distinct decision regions is a standard query-placement goal in model extraction \cite{tramer2016stealing,pal2020activethief}. Every new Autolycus query is a perturbation of an earlier seed or query. Under a limited budget, repeated local perturbations may concentrate queries around the initial seeds and leave distant regions unexplored. Phase~1 introduces non-local proposals to broaden input-space coverage.

\paragraph{W2: the traversal never searches for a boundary}
Boundary-adjacent queries can constrain a surrogate more than queries from a class interior \cite{tramer2016stealing,pal2020activethief}. Autolycus steps $\pm\varepsilon$ from the current sample along selected features, so it crosses a boundary only incidentally. It does not take two oppositely predicted samples and search the segment between them. Phase~2 adds that search.

\paragraph{W3: the combined attack obscures where its gain comes from.}
Autolycus evaluates query traversal, explanation-guided feature selection, and LIME bin-edge placement as one combined attack rather than as separate components. Any observed improvement may therefore be attributed too broadly to explanations, without showing whether it comes from the attribution ranking, the bin edges, or the traversal itself. Moreover, even if an explanation component helps, the combined result does not show whether it provides information specific to the service or information that an attacker can reconstruct from its own data. RQ1--RQ3 separate these possibilities

\paragraph{Withholding bin edges as a mitigation}
A service could return attribution weights while withholding LIME bin edges. This prevents the base Autolycus traversal from using the feature values returned by LIME, but it is protective only if the attacker cannot construct a useful substitute grid. Whether an attacker can construct a useful substitute grid depends on how the discretizer generates its grid. LIME's default \texttt{QuartileDiscretizer} places cuts at the $25$th, $50$th, and $75$th percentiles of the background data, so auxiliary feature values can be used to estimate a similar grid. LIME's entropy discretizer instead fits label-dependent trees. In our setting, the attacker uses the predictions returned for its initial seed queries as those labels. RQ3 tests both conditions: the default quartile grid and an entropy-grid sensitivity check using target predictions on the initial inputs. \sys\ uses an auxiliary grid together with the coverage and boundary-search components motivated by W1 and W2.

\section{\sys}
\label{sec:method}
\sys\ addresses the three limitations in \S\ref{sec:analysis} with three phases. Before the attack, the attacker fits LIME's default \texttt{QuartileDiscretizer} to its auxiliary data, obtaining the attacker-side grid $\mathcal{G}_A$. At run time, \sys\ queries only the target's prediction API. It does not call the explanation API. Phase~1 creates coverage beyond the seed neighbourhoods (W1). Phase~2 searches between oppositely predicted samples (W2). Phase~3 expands the resulting pool by perturbing randomly selected features from bin edges in the auxiliary grid (W3).

Diverse query generation and bisection are established extraction techniques~\cite{tramer2016stealing,papernot2017practical,pal2020activethief}. Our contribution is to combine them with an auxiliary grid after testing whether that grid can replace the bin edges used by the base traversal.

\begin{figure*}[t]
\centering
\includegraphics[width=0.9\textwidth]{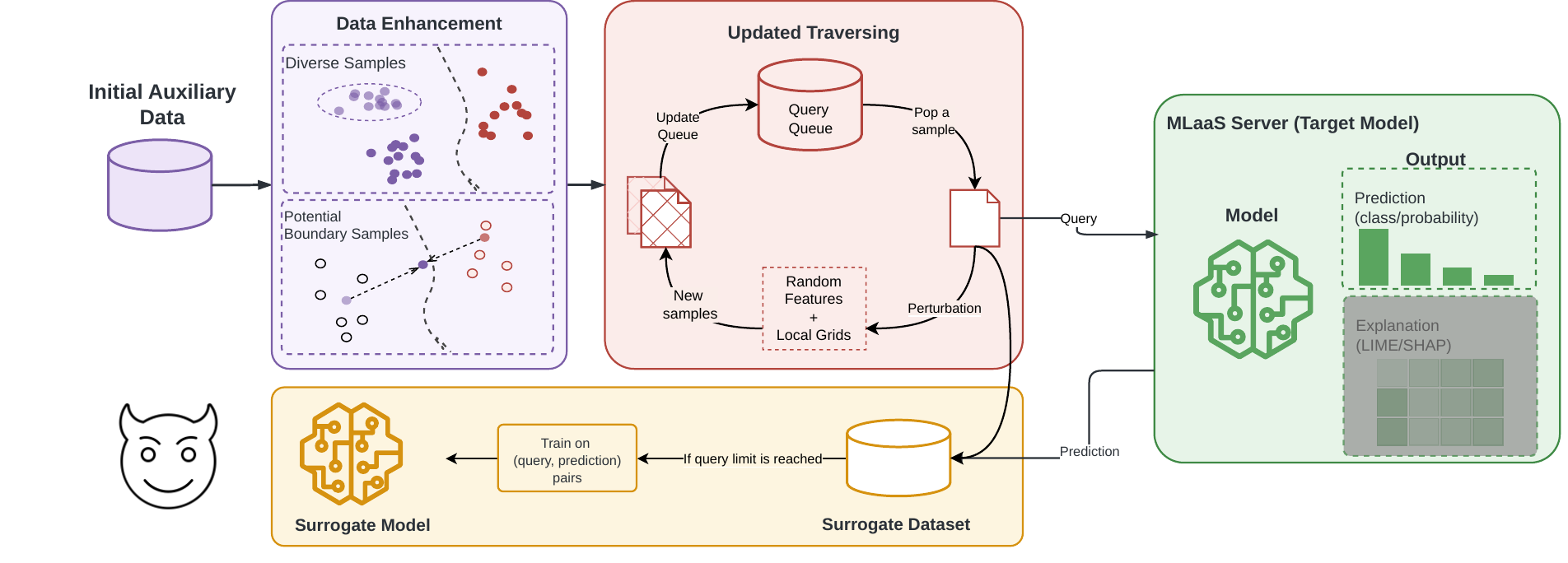}
\caption{The attacker fits a grid on auxiliary data, then uses diverse generation, boundary search, and self-grid traversal to construct a surrogate dataset. It queries the prediction API but does not query the explanation API, the greyed-out explanation endpoint is shown for context only}
\label{fig:pipeline}
\end{figure*}

Figure~\ref{fig:pipeline} sketches the pipeline. Like the base traversal, \sys\ expands each queried sample by selecting $k$ features and creating separate children that change one selected feature at a time. It selects the features at random rather than from an explanation ranking, and obtains their local bin edges from the attacker-side grid rather than from a LIME response. The per-feature offset $\varepsilon_i$ is set before the attack from the feature scale, independently of the explanation and target-model outputs. Algorithm~\ref{alg:attack} summarises the three phases.

\begin{algorithm}[t]
\caption{\sys: explanation-free model extraction. Parameter values are given in
\S\ref{subsec:diverse}--\ref{subsec:traversal}.}
\label{alg:attack}
\begin{algorithmic}[1]
\Require target prediction API $f$, auxiliary data $\mathcal{D}_A$, seeds $\mathcal{D}_0$, step sizes $\{\varepsilon_i\}$, budget $Q$
\Ensure surrogate $\hat{f}$
\State $\mathcal{G}_A \gets \textsc{QuartileDiscretizer}(\mathcal{D}_A)$ \Comment{attacker-side grid}
\State $\mathcal{S}, \mathcal{T} \gets \mathcal{D}_0$
       \Comment{traversal queue, labelled set}
\Statex \textit{// Phase 1: diverse query generation}
\State $\mathcal{D} \gets
       \{x \in \textsc{DiverseGen}(\mathcal{S}) \,:\, \max \hat{p}(x) > \tau\}$
       \Comment{retained diverse samples}
\Statex \textit{// Phase 2: boundary search}
\State $\mathcal{C} \gets \emptyset$ \Comment{boundary-search samples}
\For{$n_{\text{mid}}$ cross-class pairs $(a,b)$ drawn from $\mathcal{S}$}
  \State $I \gets \textsc{RandomFeatures}(5)$
  \State $c \gets \textsc{MakeChild}(a,b,I,\mathcal{G}_A)$
  \State $(c, M) \gets \textsc{Bisect}(c, b)$ \Comment{$8$ steps, midpoints $M$}
  \State $\mathcal{C} \gets \mathcal{C} \cup \{c\}$;\quad $\mathcal{T} \gets \mathcal{T} \cup M$
\EndFor
\State $\mathcal{S} \gets \mathcal{S} \cup \mathcal{D} \cup \mathcal{C}$
\Statex \textit{// Phase 3: self-grid traversal}
\While{$\mathcal{S} \neq \emptyset$ \textbf{and} budget and per-class quotas remain}
  \State $x \gets \textsc{Pop}(\mathcal{S})$;\quad $\mathcal{T} \gets \mathcal{T} \cup \{(x, f(x))\}$
  \State $I \gets \textsc{RandomFeatures}(k)$
  \ForAll{$i \in I$}
     \State $g_i \gets \textsc{GridBase}(x_i,\mathcal{G}_A)$, or $x_i$ if undefined
     \State add $x$ with coordinate $i$ set to $g_i \pm \varepsilon_i$, if valid and new
  \EndFor
\EndWhile
\State \Return $\hat{f} \gets \textsc{Train}(\mathcal{T})$
\end{algorithmic}
\end{algorithm}

\subsection{Phase 1: Diverse Query Generation}
\label{subsec:diverse}
Autolycus expands a small per-class seed set by local perturbation, so regions far from every seed may never be visited. 
Phase~1 is a coverage heuristic that proposes inputs by either combining feature values from two current samples (crossover) or modifying feature values of one sample (mutation). To limit distance from the original seed data, it discards proposals whose nearest original seed is farther than the $90$th percentile of seed-to-seed nearest-neighbour distances. Among the remaining proposals, it selects the one farthest from the current pool and repeats this procedure. Thus, the selected inputs are different from those already available while remaining close to the seed data. The target is queried for each selected input, and it is added to the traversal pool only when the target's maximum predicted probability exceeds $\tau=0.8$. Phase~1 uses no explanation information.

\subsection{Phase 2: Boundary Search}
\label{subsec:boundary}

Phase~2 searches for samples near a change in the target prediction. We enumerate pairs of initial seed samples with different target-predicted labels, shuffle them, and draw $n_{\text{mid}}$ pairs. For each pair $(a,b)$, \sys\ selects $5$ features at random and constructs an initial child using $a$ as an anchor. For a selected continuous feature, it sets the value to the edge in $\mathcal{G}_A$ that lies in the direction of $b$. For every other continuous feature, it samples a value uniformly between the values in $a$ and $b$. Categorical features are copied at random from either parent. It then performs a fixed-depth search between the child and $b$. At each step, it queries an intermediate point and uses the returned label to choose the segment explored next. Repeating this eight times produces eight midpoint queries intended to concentrate samples near a change in the target prediction. We fix the depth at eight in all experiments to balance this refinement against its query cost. The resulting midpoint queries are added to $\mathcal{T}$. We use up to $n_{\text{mid}}=10$ pairs and reduce this number at small budgets to reserve queries for Phase~3.

\subsection{Phase 3: Self-Grid Traversal}
\label{subsec:traversal}

Phase~3 turns each queried sample into candidates for further queries. For each
sample $x$, \sys\ selects a fixed number $k$ of features at random. Each
selected continuous feature $i$ produces two candidate children. They retain
all other feature values from $x$ and set feature $i$ just below or above its
local edge in $\mathcal{G}_A$, offset by $\varepsilon_i$. 

Before a candidate enters the queue, \sys\ enforces the feature domains: it
discards out-of-range categorical values and clips continuous values to their
observed bounds. It also discards candidates that duplicate an already queried
or queued sample. The remaining children enter the queue until the budget is
exhausted or the per-class quotas are met. This phase uses neither explanation
weights nor explanation calls.

\section{Experiments}
\label{sec:experiments}
\subsection{Setup}
\label{subsec:setup}
We evaluate on six public tabular datasets (Table~\ref{tab:datasets}) covering continuous, categorical, and mixed feature schemas, against five target families: logistic regression (LR), naive Bayes (NB), $k$-NN, decision tree (DT) and random forest (RF). We use the dataset versions and preprocessing reported for Autolycus, the explanation-guided extraction attack that we analyse throughout. In particular, for crop, five complete classes are removed from the label-sorted $2{,}200$-row source data, leaving $1{,}700$ rows and $17$ balanced classes. All five appear in every experiment: the channel ablations of \S\ref{subsec:e2} and \S\ref{subsec:e3}, the self-computability test of \S\ref{subsec:e5}, in the adversary comparison of \S\ref{subsec:ladder} and in the budget sweep. We scope to tabular data throughout, following the attack we analyse~\cite{oksuz2024autolycus}.  Following the Autolycus convention, the LIME attacker uses $n{=}1$ seed per class and explores $k{=}3$ top features, and the SHAP attacker uses $n{=}5$ and $k{=}3$. Query budgets are per dataset (Table~\ref{tab:datasets}). We inherit the baseline's query accounting unchanged: the explanation is bundled with the prediction and not charged separately, a candidate already queried or queued is not resubmitted, and fidelity is reported against a fixed maximum query budget. Phases~1 and~2 both consume budget. Every diverse candidate is queried, including those the confidence filter then discards, and every bisection midpoint is queried and retained as training data. A budget-scaling rule caps that combined overhead so it cannot starve Phase~3 at small $Q$. All compared configurations receive the same maximum budget.

Unless a setting is varied in a named ablation or sweep, we fix it across datasets, target families, and target instantiations. For \sys, Phase~1 uses $\tau=0.8$ and the $90$th percentile nearest-seed-distance filter. It generates at most $\min(\lfloor0.15Q\rfloor,15)$ diverse candidates. Phase~2 uses at most $\min(10,\max(2,\lfloor0.4Q/8\rfloor))$ cross-class pairs and eight bisection steps per pair. We selected the diverse-generation cap of $15$ once in a global sensitivity sweep and then held it fixed rather than tuning it per target or split.

Compared configurations always run on paired seed sets (identical draws, identical RNG state), and surrogate fidelity (Eq.~\ref{eq:sim}) is aggregated by the mean over refits. Aggregate contrasts include \(95\%\) percentile-bootstrap intervals over dataset--model cell means. For per-cell directional effects, we use paired Wilcoxon signed-rank tests and control the false-discovery rate at \(5\%\) with Benjamini--Hochberg adjustment within each comparison family. Complete intervals and adjusted \(p\)-values are included in the artifact.

\paragraph{Protocol} We deviate from the original attack's reporting in two ways, both toward more conservative estimates. First, the original trains $R$ surrogate refits for each attack configuration and reports the single one most similar to the target, a quantity measured on the evaluation set, so the selection is made using the reported metric. We instead report the mean over the $R$ refits, which avoids selection on the evaluation metric. ($R$ denotes surrogate refits and is unrelated to the top-$k$ feature budget $k$, and we use $R{=}20$ for DT, $R{=}10$ for RF, and $R{=}1$ otherwise.)
Secondly, the original varies only the attacker's seed sets against a single fitted target, so its reported spread reflects seed noise rather than variation across targets. For the ablations (\S\ref{subsec:e2},~\S\ref{subsec:e3}) we therefore draw $10$ independent train/test splits and retrain the target on each, pairing all configurations within a split, so each observation is an independent target instantiation. We retain the original single-split protocol only for the decomposition of Appendix~\ref{app:decomp} and the sweeps of Appendix~\ref{app:matched} and Appendix~\ref{app:sweeps}, so that those numbers stay comparable to the published baseline. Every comparison in the main text uses the multi-split protocol. 

\begin{table}[t]
\centering
\caption{Datasets. cont. counts continuous features and $Q$ is the query budget. Train, evaluation
and auxiliary partitions follow the $75\!:\!15\!:\!10$ split of the attack we analyse.}
\label{tab:datasets}
\begin{tabular}{lrrrrr}
\toprule
 & $n$ & feat. & cont. & cls. & $Q$ \\
\midrule
crop      & $1{,}700$  & 7  & 7  & 17 & $1000$ \\
adult     & $32{,}561$ & 11 & 3  & 2  & $1000$ \\
breast    & $569$      & 30 & 30 & 2  & $100$  \\
nursery   & $12{,}960$ & 8  & 0  & 3  & $1000$ \\
mushroom  & $8{,}123$  & 21 & 0  & 2  & $1000$ \\
pendigits & $10{,}992$ & 16 & 16 & 10 & $1000$ \\
\bottomrule
\end{tabular}
\end{table}

\subsection{RQ1: Does the attribution ranking help?}
\label{subsec:e2}
We replace explanation-ranked feature selection with a random-$k$ choice while holding the rest of the base traversal fixed. If an attribution ranking provides a reliable advantage, random selection should reduce fidelity.

For \textsc{Shap}, ranked selection has no overall advantage: the mean difference over the $30$ cells is $-0.16$pp (95\% interval $[-1.77,+1.35]$. Table~\ref{tab:shap-attr}). The two largest positive effects occur for TreeSHAP on the decision-tree and random-forest targets trained on mushroom (DT $+9.8$pp and RF $+6.9$pp). However, SHAP-guided selection reduces fidelity for mushroom logistic regression, which uses KernelSHAP ($-11.2$pp), and for several nursery targets. For \textsc{Lime}, where both configurations retain the bin edge, the mean difference is $-0.64$pp (95\% interval $[-2.53,+1.18]$. Table~\ref{tab:lime-attr}) and the median is $-1.0$pp. Individual LIME effects occur in both directions, ranging from $-15.4$pp to $+7.7$pp. These experiments do not establish a general rule for predicting which targets benefit from ranking.

Thus, attribution-ranked top-$k$ feature selection has no overall mean advantage in this base traversal, although it improves fidelity for some targets. This experiment does not test other uses of attribution values, such as choosing perturbation sizes or allocating queries. RQ2 tests the separate contribution of LIME bin edges.

\begin{table}[t]
\centering
\caption{RQ1: SHAP-ranked versus random feature selection. Mean paired fidelity difference (pp). Positive values mean SHAP-ranked selection achieves higher fidelity.}
\label{tab:shap-attr}
\setlength{\tabcolsep}{3pt}
\begin{tabular}{lrrrrrr}
\toprule
 & crop & pendigits & breast & adult & nursery & mushroom \\
\midrule
LR  & $-5.5$ & $-1.9$  & $+1.2$ & $-0.0$ & $-10.2$ & $-11.2$ \\
NB  & $-1.0$ & $+0.6$  & $+1.4$ & $+2.5$ & $-9.1$  & $+3.8$ \\
KNN & $+1.0$ & $+1.0$  & $-0.3$ & $+1.0$ & $-4.6$  & $+3.6$ \\
DT  & $+1.3$ & $+2.7$  & $+2.3$ & $+0.7$ & $-1.1$  & $+9.8$ \\
RF  & $+0.4$ & $-1.5$  & $+0.3$ & $+0.7$ & $+0.7$  & $+6.9$ \\
\bottomrule
\end{tabular}
\end{table}

\begin{table}[t]
\centering
\caption{RQ1: LIME-ranked versus random feature selection. Mean paired fidelity difference (pp).
Bin edges are available to both configurations. Positive values mean LIME-ranked selection achieves higher fidelity.}
\label{tab:lime-attr}
\setlength{\tabcolsep}{3pt}
\begin{tabular}{lrrrrrr}
\toprule
 & crop & pendigits & breast & adult & nursery & mushroom \\
\midrule
LR  & $-3.8$ & $+0.1$ & $+6.0$ & $-4.5$ & $-0.2$ & $-2.9$ \\
NB  & $+6.8$ & $-1.4$ & $+5.2$ & $-1.8$ & $-6.0$ & $-15.4$ \\
KNN & $+4.4$ & $-0.6$ & $+4.0$ & $-0.3$ & $+5.7$ & $+6.1$ \\
DT  & $-2.2$ & $-5.7$ & $-2.7$ & $-6.0$ & $+1.5$ & $+5.3$ \\
RF  & $-4.0$ & $-6.9$ & $-5.6$ & $-3.7$ & $+1.5$ & $+7.7$ \\
\bottomrule
\end{tabular}
\end{table}

\subsection{RQ2: Do LIME bin edges help?}
\label{subsec:e3}
Like RQ1, this experiment uses the base traversal, without \sys's diverse generation or boundary search. Following the baseline, both configurations perturb the same top-$k$ features selected by LIME with a fixed unit step. The bin-edge configuration produces $l_i-1$ and $u_i+1$, where $l_i$ and $u_i$ are LIME's returned bin edges. The control produces $x_i-1$ and $x_i+1$ from the current feature value $x_i$. Thus, the configurations differ only in the coordinate from which the baseline's unit step begins.

\begin{table}[t]
\centering
\caption{RQ2: Effect of LIME bin edges. Mean paired fidelity difference (pp). Positive values mean bin-edge traversal achieves higher fidelity.}
\label{tab:lime-threshold}
\setlength{\tabcolsep}{3pt}
\begin{tabular}{lrrrrrr}
\toprule
 & crop & pendigits & breast & adult & nursery & mushroom \\
\midrule
LR  & $+40.9$ & $+18.6$ & $-10.2$ & $-1.8$ & $-0.8$ & $-8.7$ \\
NB  & $+24.9$ & $+7.9$  & $+11.2$ & $+0.0$ & $+5.0$ & $-0.2$ \\
KNN & $+15.2$ & $+0.4$  & $-0.2$  & $-0.3$ & $+1.0$ & $+13.9$ \\
DT  & $+38.0$ & $+10.1$ & $+4.8$  & $+2.6$ & $+0.6$ & $+2.1$ \\
RF  & $+22.7$ & $+0.7$  & $+1.1$  & $+0.9$ & $+0.5$ & $-0.8$ \\
\bottomrule
\end{tabular}
\end{table}

Across all $30$ cells, bin-edge stepping improves fidelity by $+6.66$pp on average (95\% interval $[+2.70,+11.14]$), and $12$ cells retain a positive directional effect after Benjamini--Hochberg adjustment. The largest gains occur on crop and pendigits (Table~\ref{tab:lime-threshold}). On crop, bin-edge stepping improves fidelity by $+15$ to $+41$pp across the five targets. On pendigits, it improves LR, NB, and DT by $+8$ to $+19$pp. Unlike RQ1, these large effects have a consistent positive direction.

The interpretation of a bin edge depends on the feature type. For a continuous feature, an edge is a numeric value at which to place a nearby query. A categorical feature has no corresponding ordered boundary. If integer-encoded categories are discretized, an apparent edge reflects the arbitrary encoding rather than a meaningful change in the feature. A separate issue arises for near-ceiling targets: when base fidelity is already high, better query placement has little room to increase measured agreement. Thus, the useful bin edges in this study occur on continuous tabular features. They are candidate query values, not decision-boundary estimates.

\subsection{RQ3: Can auxiliary data reproduce the LIME grid?}
\label{subsec:e5}
RQ2 experiments show that LIME bin edges can help the base traversal. RQ3 asks whether an attacker can construct an equivalent grid from auxiliary data rather than obtain those edges from LIME. LIME's default \texttt{QuartileDiscretizer} places bin edges at the $25$th, $50$th, and $75$th percentiles of its background data. Those percentiles depend on the data used to fit the discretizer, not on the target model or its labels.

RQ3 focuses on the $12$ dataset-model pairs in which RQ2 found a positive bin-edge effect that passed the paired robustness check. The question here is whether an attacker can reproduce a useful grid, not whether bin edges help every target.

Each paired trial has three otherwise identical base-traversal configurations. The traversal, seed inputs, query budget, and random state are fixed, so only the grid source varies. \texttt{nothresh} uses no bin edge. \texttt{tgt} uses a LIME grid fitted on the target's training features $X_{\text{train}}$, representing the grid from which the service returns bin edges. \texttt{aux} uses a grid fitted on the attacker's auxiliary data $D_A$. We report each grid's fidelity gain over \texttt{nothresh}. The target advantage, $\texttt{tgt}-\texttt{aux}$, directly compares the two grid sources: a value near zero means that the auxiliary grid is an effective substitute.

\begin{table}[t]
\centering
\caption{RQ3: Target and auxiliary LIME grids. Fidelity gain over no-bin-edge traversal (pp).
Target advantage $=\textsc{tgt}-\textsc{aux}$, so negative values favour the auxiliary grid.}
\label{tab:auxdisc}
\begin{tabular}{llrrr}
\toprule
dataset & model & tgt & aux & target adv. \\
\midrule
crop      & LR  & $+40.9$ & $+41.9$ & $-0.9$ \\
crop      & DT  & $+38.0$ & $+36.8$ & $+1.1$ \\
crop      & NB  & $+24.9$ & $+24.1$ & $+0.9$ \\
crop      & RF  & $+22.7$ & $+22.4$ & $+0.3$ \\
pendigits & LR  & $+18.6$ & $+20.1$ & $-1.5$ \\
crop      & KNN & $+15.2$ & $+17.7$ & $-2.5$ \\
mushroom  & KNN & $+13.9$ & $+15.8$ & $-1.9$ \\
breast    & NB  & $+11.2$ & $+11.9$ & $-0.7$ \\
pendigits & DT  & $+10.1$ & $+13.5$ & $-3.4$ \\
pendigits & NB  & $+7.9$  & $+7.8$  & $+0.2$ \\
nursery   & NB  & $+5.0$  & $+4.7$  & $+0.3$ \\
nursery   & KNN & $+1.0$  & $+1.3$  & $-0.3$ \\
\bottomrule
\end{tabular}
\end{table}

Across the twelve cells in Table~\ref{tab:auxdisc}, the auxiliary grid reproduces the gain. The mean target-grid advantage is $-0.71$pp (95\% interval $[-1.51,+0.01]$) and ranges from $-3.4$ to $+1.1$pp, while the bin-edge effect itself reaches $+40.9$pp. On crop/KNN, the auxiliary grid is $2.5$pp higher than the target grid. The target grid does not show a consistent advantage in this controlled comparison.

\paragraph{Grid-data sensitivity} The comparison above fits the auxiliary grid on the full baseline-compatible partition. We then fit it using only the $n$ inputs per class that initialize the attack, leaving the controlled traversal unchanged. Table~\ref{tab:auxsize-main} shows selected cases from the full sweep in Appendix~\ref{app:auxsize}. On crop, one input per class already matches the target-grid gain for LR and DT. Pendigits is more demanding: for LR, the auxiliary gain rises from $+12.9$pp with one input per class to $+15.4$pp with two, against $+18.6$pp for the target grid. Thus, the full $10\%$ partition is not required to fit a useful grid in these tests. The analysis retains the initial small seed pool and does not make the entire attack data-free directly.

\begin{table}[t]
\centering
\setlength{\tabcolsep}{5pt}
\caption{RQ3: Quartile grids fitted on a small initial pool. Fidelity gain over no-bin-edge traversal (pp). Target uses target training features. $n$ is the number of initial inputs per class used to fit the auxiliary grid.  Full results are in Appendix~\ref{app:auxsize}.}
\label{tab:auxsize-main}
\begin{tabular}{llrrr}
\toprule
dataset & model & target & $n{=}1$ & $n{=}2$ \\
\midrule
crop      & LR & $+40.9$ & $+40.7$ & $+43.9$ \\
crop      & DT & $+38.0$ & $+38.2$ & $+38.2$ \\
pendigits & LR & $+18.6$ & $+12.9$ & $+15.4$ \\
pendigits & DT & $+10.1$ & $+9.3$  & $+9.3$  \\
\bottomrule
\end{tabular}
\end{table}

\subsubsection{Class-support restriction.} The preceding analyses draw seeds and grid-fitting data
from the same auxiliary partition. We next restrict the attacker to a subset of the auxiliary
classes, while the target and evaluation set retain every class. We test all five crop targets and
the three pendigits targets with a positive RQ2 effect. A moderate restriction retains $11$ of $17$
crop classes and $7$ of $10$ digit classes. A severe restriction retains $6$ and $4$, respectively.
For each restriction, the target-grid and attacker-grid configurations use the same restricted seeds,
budget, traversal, and random state. Their difference therefore measures the value of the service
grid conditional on the attacker's available classes.

\begin{table}[t]
\centering
\caption{RQ3: Class-support sensitivity. Target-grid advantage (pp) with matched restricted seeds.}
\label{tab:auxshift}
\setlength{\tabcolsep}{6pt}
\begin{tabular}{lrr}
\toprule
restriction & target advantage & cells favouring target \\
\midrule
moderate & $+1.18$ & $6/8$ \\
severe   & $+5.41$ & $8/8$ \\
\bottomrule
\end{tabular}
\end{table}

Under the moderate restriction, the target grid has only a small mean advantage. Under the severe
restriction, it is higher in all eight evaluated cells, by $+5.41$pp on average. The attacker-side
grid remains useful even in this setting: relative to the same restricted attack without bin edges,
it improves fidelity by $+22.3$pp under the moderate restriction and $+20.4$pp under the severe
one. Thus, a restricted attacker can still benefit from self-computed bin edges, but under severe
class-support restriction they no longer fully substitute for the service grid. Appendix~\ref{app:auxshift}
gives the per-cell results and separates the effects of restricting the grid, the seeds, and both.

\subsubsection{Entropy discretization.} The default quartile discretizer uses no labels. We therefore also test LIME's label-dependent entropy discretizer, which fits a small decision tree for each feature. The attacker uses the cached target predictions on two initial inputs per class as the discretizer labels. This adds no separate query type because the same predictions initialize the attack. Table~\ref{tab:entropy-two} reports the seven configurations in which entropy bin edges improve the base traversal. It gives each grid's gain over no-bin-edge traversal and the target advantage, defined as target minus auxiliary. The auxiliary entropy grid retains these gains. Its difference from the target grid ranges from $-2.7$pp to $+3.7$pp, with no consistent direction. Thus, even for this label-dependent discretizer, two initial inputs per class are sufficient for the reported comparisons. Entropy-tree tie breaking is random, so results at other pool sizes need not vary monotonically.

\begin{table}[t]
\centering
\setlength{\tabcolsep}{4pt}
\caption{RQ3: Entropy-grid results with two initial inputs per class. Fidelity gain over
no-bin-edge traversal (pp).}
\label{tab:entropy-two}
\begin{tabular}{llrrr}
\toprule
dataset & model & target & auxiliary & target adv. \\
\midrule
crop      & LR  & $+20.3$ & $+22.9$ & $-2.6$ \\
crop      & NB  & $+19.1$ & $+17.6$ & $+1.5$ \\
crop      & KNN & $+13.5$ & $+13.1$ & $+0.4$ \\
crop      & DT  & $+33.3$ & $+29.6$ & $+3.7$ \\
crop      & RF  & $+13.7$ & $+11.5$ & $+2.2$ \\
pendigits & LR  & $+7.3$  & $+10.0$ & $-2.7$ \\
pendigits & DT  & $+7.4$  & $+5.6$  & $+1.8$ \\
\bottomrule
\end{tabular}
\end{table}

\begin{table*}[t]
\centering
\caption{Evaluation of \sys. Mean fidelity (\%) and pairwise contrasts (pp). \textsc{sg}$-$\textsc{aut}
compares the auxiliary grid and random feature selection with the target explanation under the base
traversal. \textsc{self-grid} and \sys\ make no explanation calls.}
\label{tab:ladder}
\begin{tabular}{llrrrrrrrr}
\toprule
& & \multicolumn{5}{c}{fidelity} & \multicolumn{3}{c}{contrast} \\
\cmidrule(lr){3-7}\cmidrule(lr){8-10}
dataset & model & \textsc{blind} & \textsc{defended} & \textsc{autolycus} & \textsc{self-grid}
& \sys\ & \textsc{sg}$-$\textsc{aut} & \sys$-$\textsc{aut} & \sys$-$\textsc{def} \\
\midrule
crop       & LR   & $38.9$ & $39.5$ & $80.4$ & $82.3$ & $91.6$ & $+1.8$ & $+11.1$ & $+52.1$ \\
crop       & NB   & $57.5$ & $56.0$ & $80.9$ & $72.7$ & $75.4$ & $-8.2$ & $-5.5$ & $+19.5$ \\
crop       & KNN  & $53.6$ & $51.4$ & $66.6$ & $61.4$ & $63.1$ & $-5.3$ & $-3.5$ & $+11.6$ \\
crop       & DT   & $40.0$ & $39.7$ & $77.6$ & $77.9$ & $89.9$ & $+0.3$ & $+12.3$ & $+50.3$ \\
crop       & RF   & $64.3$ & $59.5$ & $82.2$ & $86.1$ & $91.3$ & $+3.9$ & $+9.2$ & $+31.9$ \\
\addlinespace
pendigits  & LR   & $57.2$ & $48.9$ & $67.5$ & $70.3$ & $86.1$ & $+2.7$ & $+18.5$ & $+37.1$ \\
pendigits  & NB   & $66.7$ & $65.2$ & $73.1$ & $72.4$ & $84.1$ & $-0.8$ & $+11.0$ & $+18.9$ \\
pendigits  & KNN  & $62.5$ & $62.5$ & $62.8$ & $64.2$ & $68.9$ & $+1.4$ & $+6.1$ & $+6.4$ \\
pendigits  & DT   & $30.7$ & $32.6$ & $42.7$ & $48.2$ & $49.6$ & $+5.5$ & $+6.9$ & $+16.9$ \\
pendigits  & RF   & $47.6$ & $50.5$ & $51.2$ & $56.7$ & $73.4$ & $+5.4$ & $+22.1$ & $+22.9$ \\
\addlinespace
breast     & LR   & $68.9$ & $72.8$ & $62.6$ & $60.7$ & $78.0$ & $-1.9$ & $+15.4$ & $+5.2$ \\
breast     & NB   & $74.1$ & $75.6$ & $86.8$ & $80.7$ & $88.9$ & $-6.1$ & $+2.1$ & $+13.3$ \\
breast     & KNN  & $69.1$ & $71.7$ & $71.4$ & $66.2$ & $73.8$ & $-5.2$ & $+2.4$ & $+2.1$ \\
breast     & DT   & $63.8$ & $64.0$ & $68.8$ & $72.2$ & $73.4$ & $+3.4$ & $+4.6$ & $+9.4$ \\
breast     & RF   & $85.8$ & $75.5$ & $76.6$ & $85.3$ & $82.5$ & $+8.7$ & $+5.9$ & $+7.0$ \\
\addlinespace
adult      & LR   & $81.6$ & $81.3$ & $79.5$ & $85.1$ & $86.4$ & $+5.6$ & $+6.9$ & $+5.1$ \\
adult      & NB   & $87.8$ & $87.6$ & $87.7$ & $89.6$ & $89.9$ & $+2.0$ & $+2.2$ & $+2.2$ \\
adult      & KNN  & $82.8$ & $84.0$ & $83.8$ & $84.2$ & $86.7$ & $+0.4$ & $+3.0$ & $+2.7$ \\
adult      & DT   & $64.4$ & $68.4$ & $71.0$ & $75.1$ & $71.2$ & $+4.1$ & $+0.2$ & $+2.8$ \\
adult      & RF   & $81.0$ & $78.9$ & $79.8$ & $83.4$ & $87.5$ & $+3.7$ & $+7.8$ & $+8.6$ \\
\addlinespace
nursery    & LR   & $98.1$ & $98.7$ & $97.9$ & $98.1$ & $97.9$ & $+0.2$ & $+0.0$ & $-0.8$ \\
nursery    & NB   & $84.0$ & $79.1$ & $84.1$ & $90.4$ & $80.9$ & $+6.3$ & $-3.2$ & $+1.7$ \\
nursery    & KNN  & $69.0$ & $80.5$ & $81.5$ & $76.5$ & $71.0$ & $-5.0$ & $-10.5$ & $-9.5$ \\
nursery    & DT   & $87.1$ & $89.9$ & $90.5$ & $88.4$ & $88.7$ & $-2.0$ & $-1.8$ & $-1.2$ \\
nursery    & RF   & $86.2$ & $90.3$ & $90.8$ & $90.7$ & $88.7$ & $-0.1$ & $-2.1$ & $-1.6$ \\
\addlinespace
mushroom   & LR   & $85.3$ & $92.3$ & $83.6$ & $82.2$ & $80.8$ & $-1.4$ & $-2.8$ & $-11.5$ \\
mushroom   & NB   & $71.3$ & $65.8$ & $65.6$ & $76.8$ & $79.7$ & $+11.1$ & $+14.0$ & $+13.8$ \\
mushroom   & KNN  & $65.5$ & $63.7$ & $77.6$ & $72.6$ & $70.4$ & $-5.1$ & $-7.2$ & $+6.7$ \\
mushroom   & DT   & $72.3$ & $81.7$ & $83.7$ & $78.6$ & $77.3$ & $-5.2$ & $-6.5$ & $-4.4$ \\
mushroom   & RF   & $66.8$ & $81.5$ & $80.7$ & $74.1$ & $75.7$ & $-6.6$ & $-5.0$ & $-5.7$ \\
\midrule
\multicolumn{2}{l}{mean} & $68.8$ & $69.6$ & $76.3$ & $76.8$ & $80.1$ & $+0.5$ & $+3.8$ & $+10.4$ \\
\bottomrule
\end{tabular}
\end{table*}

\subsection{Evaluation of \sys\ without explanation calls}
\label{subsec:ladder}
RQ1--RQ3 isolate information obtained from an explanation and test whether the target grid is
replaceable. We now evaluate \sys, which combines an auxiliary grid with diverse generation and
boundary search. Unlike RQ3, this comparison evaluates attack performance rather than
service-specific disclosure. Table~\ref{tab:ladder} compares five configurations under the same maximum budget,
auxiliary pool, per-split seeds, surrogate family, and evaluation set:

\begin{itemize}
  \item \textsc{blind}: the base traversal with random feature selection and no grid. This is the prediction-only floor.
  \item \textsc{defended}: the base traversal with the service's attribution ranking but without bin edges.
  \item \textsc{autolycus}: the base traversal using the service's attribution ranking and bin edges.
  \item \textsc{self-grid}: the base traversal with random feature selection and an auxiliary grid, without explanation calls.
  \item \sys: \textsc{self-grid} plus diverse generation and boundary search, without explanation calls.
\end{itemize}

The \textsc{self-grid}$-\textsc{autolycus}$ contrast is an end-to-end check: it removes both the service ranking and service bin edges while keeping the base traversal. It averages $+0.46$ percentage points over all $30$ cells (95\% interval $[-1.25,+2.19]$), with differences in both directions. This result is consistent with the controlled RQ3 test. It is not a replacement for it. The three prediction-only configurations---\textsc{blind}, \textsc{self-grid}, and \sys---make no explanation calls. Their grids are fitted from auxiliary data before the attack begins.

\paragraph{What the additional query placement contributes} The contrast \sys$-\textsc{self-grid}$ holds the auxiliary grid fixed and isolates diverse generation plus boundary search. It averages $+3.3$ percentage points across all cells. On the four datasets with continuous features, it is $+5.9$ percentage points. On the two categorical datasets, it is $-1.7$ percentage points. Across all $30$ cells, \sys$-\textsc{autolycus}$ averages $+3.79$pp (95\% interval $[+0.99,+6.65]$), with $12$ positive directional effects after Benjamini--Hochberg adjustment. On continuous-feature datasets, the contrast is $+6.93$pp (95\% interval $[+4.04,+9.92]$); on categorical datasets, it is $-2.5$pp. The additional phases use grid values to place continuous-feature queries, so the categorical results are outside the method's intended scope. For example, on nursery/KNN, Autolycus reaches $81.5\%$ and \sys\ reaches $71.0\%$.

The component decomposition in Appendix~\ref{app:decomp} separates the two additions. Their average contributions are similar ($+1.6$ percentage points for boundary search and $+1.8$ for diverse generation), but the pattern depends on the target: diverse generation contributes most for naive Bayes, while boundary search contributes most for logistic regression and random forests.

\begin{figure*}[t]
\centering
\includegraphics[width=\textwidth]{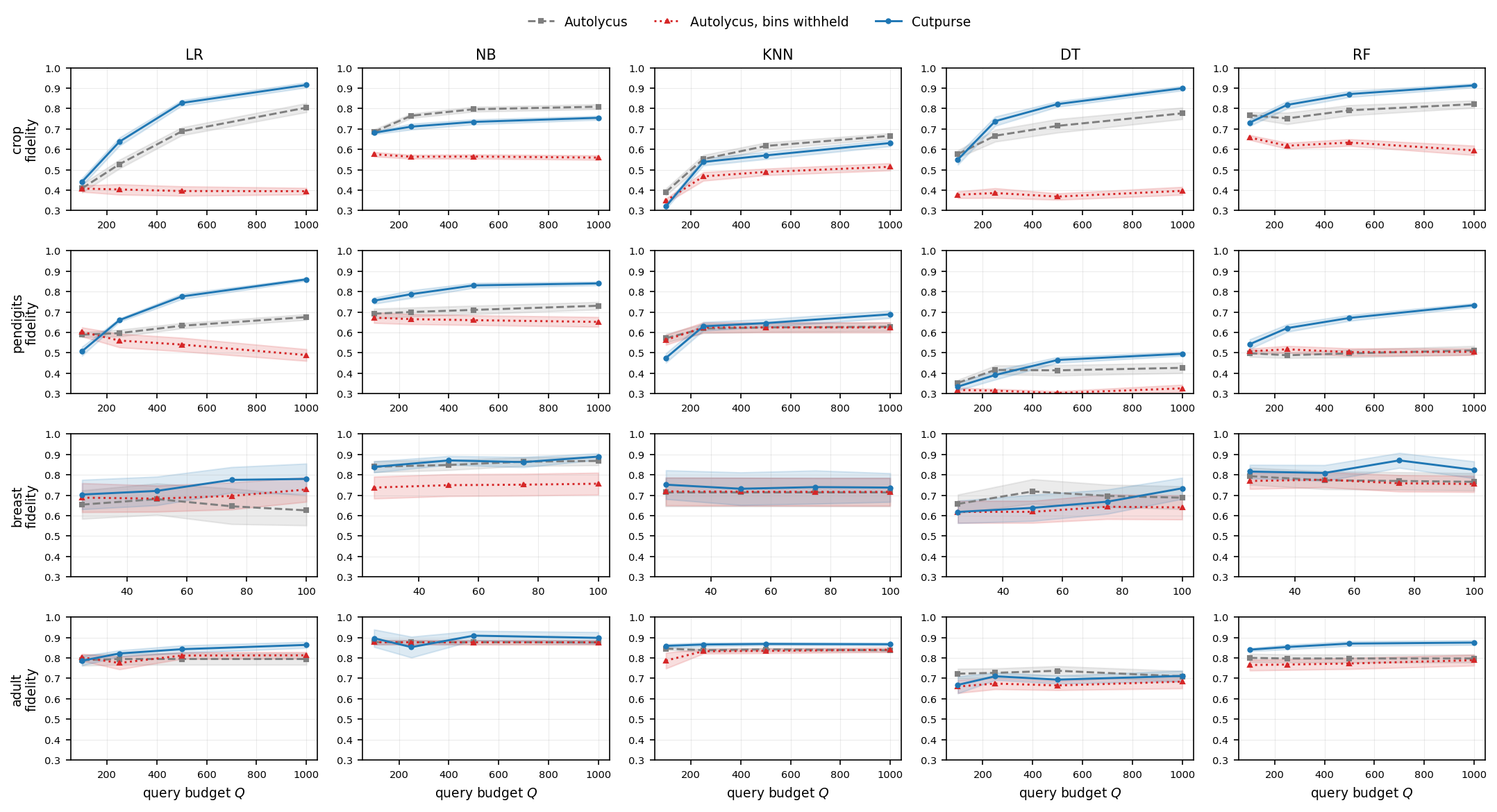}
\caption{Fidelity vs.\ query budget for Autolycus (grey, dashed), Autolycus with bin edges
withheld (red, dotted) and \sys\ (blue, solid), per dataset (rows) and target (columns), for the four
datasets with continuous features. Band is the standard error over ten independent target
instantiations. Same splits, seed sets and attack RNG as Table~\ref{tab:ladder}, whose cells are the
rightmost point of each curve.}
\label{fig:defense-sweep}
\end{figure*}

\paragraph{Effect of withholding bin edges} The defended service returns attribution weights but omits LIME bin edges. This reduces Autolycus's fidelity on some continuous-feature targets: crop/LR falls from $80.4\%$ to $39.5\%$, and crop/DT from $77.6\%$ to $39.7\%$. Across the continuous-feature datasets, replacing the defended configuration with \textsc{self-grid} raises fidelity by $10.5$ percentage points on average. Adding diverse generation and boundary search raises it by a further $5.9$ percentage points. Across all $30$ cells, \sys\ exceeds the defended configuration by $+10.45$pp (95\% interval $[+5.28,+16.26]$), with $15$ positive directional effects after Benjamini--Hochberg adjustment; on continuous-feature targets, the contrast is $+16.29$pp (95\% interval $[+10.19,+23.27]$). The first contrast combines self-supplied bin edges with random feature selection, while RQ1 finds no systematic benefit from the removed ranking for feature selection in the base traversal.

Appendix~\ref{app:steps} matches \sys\ and Autolycus on either perturbation-step rule. The \sys\ advantage remains positive under both matched comparisons.

Figure~\ref{fig:defense-sweep} shows that this difference can grow with the budget. On crop/DT, the Autolycus--defended gap rises from $19.9$ to $38.0$ percentage points. On crop/NB, it rises from $11.0$ to $24.9$. At $Q{=}100$, however, \sys\ can trail both baselines because it allocates up to $40$ queries to bisection and up to $15$ to diverse generation. The curves cross by $Q{=}250$ in the reported examples. The effect is not universal: on pendigits/KNN, withholding the bin edges changes little at every budget. These results assume in-domain auxiliary inputs. RQ3 and the grid-data sensitivity analysis identify how little of that data is needed to fit the grids tested here.

\section{Discussion and Limitations}
\label{sec:discussion}
\paragraph{Implication for default LIME} In this setting, LIME bin edges can be useful to an extraction attack, but RQ3 shows that their source need not be the service. The default quartile edges are statistics of the explainer background data. An auxiliary grid reproduces their effect in our controlled comparisons. Withholding bin edges therefore affects Autolycus, but it does not lower the performance of \sys\ under our auxiliary-data condition. This is not a claim about all explanation methods or all LIME configurations.

\paragraph{Scope.} Cutpurse is intended for tabular targets with genuine continuous features. It uses an auxiliary grid to place single-feature queries, so it is not competitive on the categorical datasets. In our LIME setup, those datasets are integer encoded and the discretizer treats the columns as continuous. Their quartile bin edges depend on an arbitrary category ordering. The TreeSHAP results provide a separate qualification to RQ1: on the mushroom decision-tree and random-forest targets, TreeSHAP rankings help feature selection in the base traversal. Ranking-based feature selection is not reliable across target families. We do not generalize these results to neural or high-dimensional targets, other explainer configurations, or label-dependent discretizers.

\paragraph{Auxiliary data and measurement limits.} The main comparisons use the $10\%$ auxiliary partition of Autolycus for baseline compatibility. The grid-data sensitivity analysis shows that fitting the grid can instead use only the initial class-balanced pool. It does not make the complete attack data-free: it retains in-domain seeds and the other fixed attack settings. Fidelity on \textsc{breast} is measured on $85$ held-out points, so a single disagreement changes it by $1.18$ percentage points. We therefore do not interpret small effects on that dataset. Finally, Phase~1's input-space diversity objective is a proxy for decision-cell coverage rather than a direct coverage measurement.

\section{Conclusion}
For the default LIME and SHAP configurations on the tabular classifiers we study, using attribution rankings for feature selection in the base Autolycus traversal provides no systematic extraction advantage. LIME bin edges can improve Autolycus by up to $41$ percentage points on continuous-feature targets. However, the controlled grid-substitution test shows no consistent advantage for the target grid over an attacker-side grid. The grid-data sensitivity analysis further shows that the full $10\%$ partition is not required to fit a useful grid in the tested cases.

\sys\ uses such an auxiliary grid, diverse query generation, and boundary search without calling an explanation API. On continuous-feature targets, it improves fidelity by $16.3$ percentage points over the defended service and by $7.0$ over Autolycus. These latter comparisons measure a stronger attack, not additional disclosure. The result is conditional on the tabular, continuous-feature, auxiliary-data setting studied here. Determining which explanation outputs remain unavailable to an adaptive adversary is an open question.

\section*{Open Science}
The artifact containing the implementation, stored experimental outputs, and scripts  will be released.

\bibliographystyle{IEEEtran}
\bibliography{sample-base}

@inproceedings{tramer2016stealing,
  title     = {Stealing Machine Learning Models via Prediction {API}s},
  author    = {Tram\`er, Florian and Zhang, Fan and Juels, Ari and Reiter, Michael K and Ristenpart, Thomas},
  booktitle = {USENIX Security},
  year      = {2016}
}

@inproceedings{orekondy2019knockoff,
  title     = {Knockoff Nets: Stealing Functionality of Black-Box Models},
  author    = {Orekondy, Tribhuvanesh and Schiele, Bernt and Fritz, Mario},
  booktitle = {CVPR},
  year      = {2019}
}

@inproceedings{jagielski2020high,
  title     = {High Accuracy and High Fidelity Extraction of Neural Networks},
  author    = {Jagielski, Matthew and Carlini, Nicholas and Berthelot, David and Kurakin, Alex and Papernot, Nicolas},
  booktitle = {USENIX Security},
  year      = {2020}
}

@inproceedings{papernot2017practical,
  title     = {Practical Black-Box Attacks Against Machine Learning},
  author    = {Papernot, Nicolas and McDaniel, Patrick and Goodfellow, Ian and Jha, Somesh and Celik, Z Berkay and Swami, Ananthram},
  booktitle = {AsiaCCS},
  year      = {2017}
}

@inproceedings{correia2018copycat,
  title     = {Copycat {CNN}: Stealing Knowledge by Persuading Confession with Random Non-Labeled Data},
  author    = {Correia-Silva, Jacson Rodrigues and Berriel, Rodrigo F and Badue, Claudine and De Souza, Alberto F and Oliveira-Santos, Thiago},
  booktitle = {IJCNN},
  year      = {2018}
}

@inproceedings{karmakar2023marich,
  title     = {Marich: A Query-efficient Distributionally Equivalent Model Extraction Attack},
  author    = {Karmakar, Pratik and Basu, Debabrota},
  booktitle = {NeurIPS},
  year      = {2023}
}

@inproceedings{pal2020activethief,
  title     = {{ACTIVETHIEF}: Model Extraction Using Active Learning and Unannotated Public Data},
  author    = {Pal, Soham and Gupta, Yash and Shukla, Aditya and Kanade, Aditya and Shevade, Shirish K and Ganapathy, Vinod},
  booktitle = {AAAI},
  year      = {2020}
}

@inproceedings{milli2019model,
  title     = {Model Reconstruction from Model Explanations},
  author    = {Milli, Smitha and Schmidt, Ludwig and Dragan, Anca D and Hardt, Moritz},
  booktitle = {FAccT},
  year      = {2019}
}

@article{oksuz2024autolycus,
  title     = {{AUTOLYCUS}: Exploiting Explainable Artificial Intelligence ({XAI})
               for Model Extraction Attacks against Interpretable Models},
  author    = {Oksuz, Abdullah Caglar and Halimi, Anisa and Ayday, Erman},
  journal   = {Proceedings on Privacy Enhancing Technologies (PoPETs)},
  volume    = {2024},
  number    = {4},
  year      = {2024},
  note      = {arXiv:2302.02162}
}

@article{yan2023explanationleaks,
  title     = {Explanation Leaks: Explanation-guided Model Extraction Attacks},
  author    = {Yan, Anli and Huang, Teng and Ke, Lishan and Liu, Xiaozhang and Chen, Qi and Dong, Changyu},
  journal   = {Information Sciences},
  volume    = {632},
  pages     = {269--284},
  year      = {2023},
  doi       = {10.1016/j.ins.2023.03.020}
}

@inproceedings{miura2024megex,
  title     = {{MEGEX}: Data-Free Model Extraction Attack against Gradient-Based Explainable {AI}},
  author    = {Miura, Takayuki and Hasegawa, Satoshi and Shibahara, Toshiki and Yanai, Naoto},
  booktitle = {ACM Workshop on Secure and Trustworthy Deep Learning Systems (SecTL)},
  year      = {2024},
  note      = {arXiv:2107.08909}
}

@article{quan2022amplification,
  title     = {On the Amplification of Security and Privacy Risks by Post-hoc
               Explanations in Machine Learning Models},
  author    = {Quan, Pengrui and Chakraborty, Supriyo and Jeyakumar, Jeya Vikranth and Srivastava, Mani},
  journal   = {arXiv:2206.14004},
  year      = {2022}
}

@article{aivodji2020model,
  title     = {Model Extraction from Counterfactual Explanations},
  author    = {A\"ivodji, Ulrich and Bolot, Alexandre and Gambs, S\'ebastien},
  journal   = {arXiv:2009.01884},
  year      = {2020}
}

@inproceedings{wang2022dualcf,
  title     = {{DualCF}: Efficient Model Extraction Attack from Counterfactual Explanations},
  author    = {Wang, Yongjie and Qian, Hangwei and Miao, Chunyan},
  booktitle = {ACM Conference on Fairness, Accountability, and Transparency (FAccT)},
  year      = {2022},
  doi       = {10.1145/3531146.3533188},
  note      = {arXiv:2205.06504}
}

@inproceedings{khouna2025tra,
  title     = {From Counterfactuals to Trees: Competitive Analysis of Model Extraction Attacks},
  author    = {Khouna, Awa and Ferry, Julien and Vidal, Thibaut},
  booktitle = {NeurIPS},
  year      = {2025},
  note      = {arXiv:2502.05325}
}

@article{otto2026linear,
  title     = {Linear Model Extraction via Factual and Counterfactual Queries},
  author    = {Otto, Daan and Kurtz, Jannis and den Hertog, Dick and Birbil, Ilker},
  journal   = {arXiv:2602.09748},
  year      = {2026}
}

@article{nguyen2024survey,
  title     = {A Survey of Privacy-Preserving Model Explanations: Privacy Risks,
               Attacks, and Countermeasures},
  author    = {Nguyen, Thanh Tam and Huynh, Thanh Trung and Ren, Zhao and
               Nguyen, Thanh Toan and Nguyen, Phi Le and Yin, Hongzhi and Nguyen, Quoc Viet Hung},
  journal   = {arXiv:2404.00673},
  year      = {2024}
}

@article{ezzeddine2026membership,
  title     = {Explanations Leak: Membership Inference with Differential Privacy
               and Active Learning Defense},
  author    = {Ezzeddine, Fatima and Zammar, Osama and Giordano, Silvia and Ayoub, Omran},
  journal   = {arXiv:2602.03611},
  year      = {2026}
}

@article{oksuz2026lomime,
  title     = {{LoMime}: Query-Efficient Membership Inference using Model Extraction
               in Label-Only Settings},
  author    = {Oksuz, Abdullah Caglar and Halimi, Anisa and Ayday, Erman},
  journal   = {arXiv:2602.18934},
  year      = {2026}
}

@inproceedings{lundberg2017shap,
  title     = {A Unified Approach to Interpreting Model Predictions},
  author    = {Lundberg, Scott M and Lee, Su-In},
  booktitle = {NeurIPS},
  year      = {2017}
}

@article{papernot2016transferability,
  title   = {Transferability in Machine Learning: from Phenomena to Black-Box Attacks Using Adversarial Samples},
  author  = {Papernot, Nicolas and McDaniel, Patrick D. and Goodfellow, Ian J.},
  journal = {arXiv preprint arXiv:1605.07277},
  year    = {2016}
}

@inproceedings{lowd2005adversarial,
  title     = {Adversarial Learning},
  author    = {Lowd, Daniel and Meek, Christopher},
  booktitle = {Proceedings of the Eleventh ACM SIGKDD International Conference on Knowledge Discovery in Data Mining},
  pages     = {641--647},
  year      = {2005},
  publisher = {Association for Computing Machinery},
  doi       = {10.1145/1081870.1081950}
}

@inproceedings{chandrasekaran2020active,
  title     = {Exploring Connections Between Active Learning and Model Extraction},
  author    = {Chandrasekaran, Varun and Chaudhuri, Kamalika and Giacomelli, Irene and Jha, Somesh and Yan, Songbai},
  booktitle = {29th USENIX Security Symposium (USENIX Security 20)},
  pages     = {1309--1326},
  year      = {2020},
  publisher = {USENIX Association},
  url       = {https://www.usenix.org/conference/usenixsecurity20/presentation/chandrasekaran}
}

@inproceedings{juuti2019prada,
  title     = {{PRADA}: Protecting Against {DNN} Model Stealing Attacks},
  author    = {Juuti, Mika and Szyller, Sebastian and Marchal, Samuel and Asokan, N.},
  booktitle = {2019 IEEE European Symposium on Security and Privacy (EuroS\&P)},
  pages     = {512--527},
  year      = {2019},
  publisher = {IEEE},
  doi       = {10.1109/EuroSP.2019.00044}
}

@inproceedings{shokri2021privacy,
  title     = {On the Privacy Risks of Model Explanations},
  author    = {Shokri, Reza and Strobel, Martin and Zick, Yair},
  booktitle = {Proceedings of the 2021 AAAI/ACM Conference on AI, Ethics, and Society},
  pages     = {231--241},
  year      = {2021},
  publisher = {Association for Computing Machinery},
  doi       = {10.1145/3461702.3462533}
}

@inproceedings{yue2021blackbox,
  title     = {Black-Box Attacks on Sequential Recommenders via Data-Free Model Extraction},
  author    = {Yue, Zhenrui and He, Zhankui and Zeng, Huimin and McAuley, Julian},
  booktitle = {Proceedings of the Fifteenth ACM Conference on Recommender Systems},
  pages     = {44--54},
  year      = {2021},
  doi       = {10.1145/3460231.3474275}
}

\appendices
\section{Component decomposition: which phase drives the gain?}
\label{app:decomp}
\begin{figure}[t]
\centering
\includegraphics[width=0.9\linewidth]{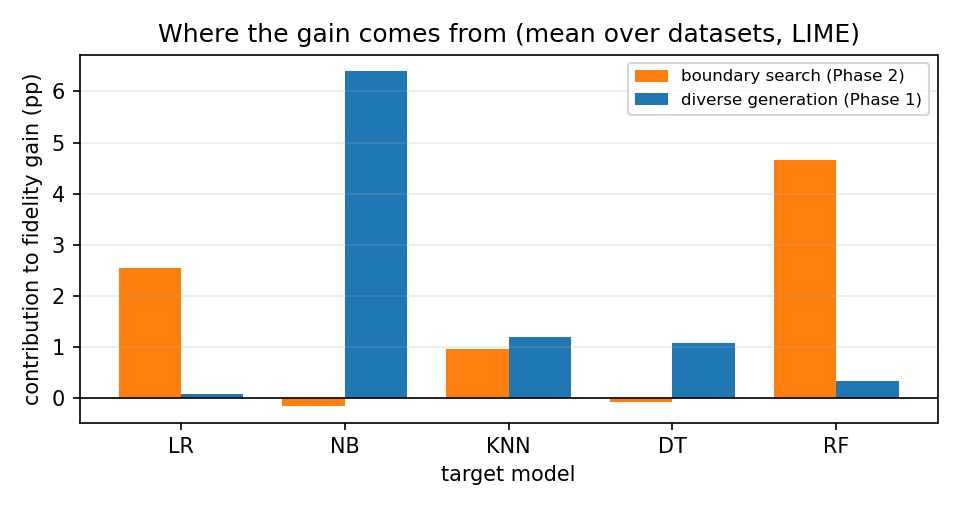}
\caption{Contribution of each component to the LIME fidelity gain over
Autolycus (percentage points, mean over datasets, top budget). Boundary search dominates for
RF/LR. Diverse generation dominates for NB.}
\label{fig:decomp}
\end{figure}

Our pipeline adds two components to the Autolycus traversal: diverse query generation (Phase~1) and
boundary search (Phase~2). We isolate their contributions with three nested configurations on the same seed sets, namely Autolycus alone (\texttt{base}), $+$boundary (Phase~2, no diverse), and $+$diverse on top, so that boundary$\,=\,$($+$boundary)$\,-\,$\texttt{base} and
diverse$\,=\,$full$\,-\,$($+$boundary). Figure~\ref{fig:decomp} reports the result.

The two components contribute about equally on average ($+1.6$ and $+1.8$pp), but which one
drives the gain is strongly target-dependent: naive Bayes benefits almost entirely from
diverse generation ($+6.4$pp), since a generative model needs distribution coverage, whereas random forest and logistic regression benefit almost entirely from boundary search
($+4.7$, $+2.5$pp), which places labelled points on the decision surface. This decomposition measures
what the added query-placement components contribute. It does not test other possible uses of
attribution magnitudes. RQ1 separately tests feature selection by attribution ranking in the base
traversal. One caveat bounds this decomposition: the boundary term as measured here bundles
Phase~2 boundary search with the Phase~3 bin-edge stepping it inherits from Autolycus, so it should
be read as the combined contribution of both rather than of bisection alone. \S\ref{subsec:e3}
isolates bin-edge stepping directly. Like the sweeps of Appendix~\ref{app:matched}, this decomposition uses the
explanation-reading configuration, so it measures what the two components add given the
target's explanation, while \S\ref{subsec:ladder} measures what they add without it.

\begin{figure*}[t]
\centering
\includegraphics[width=\textwidth]{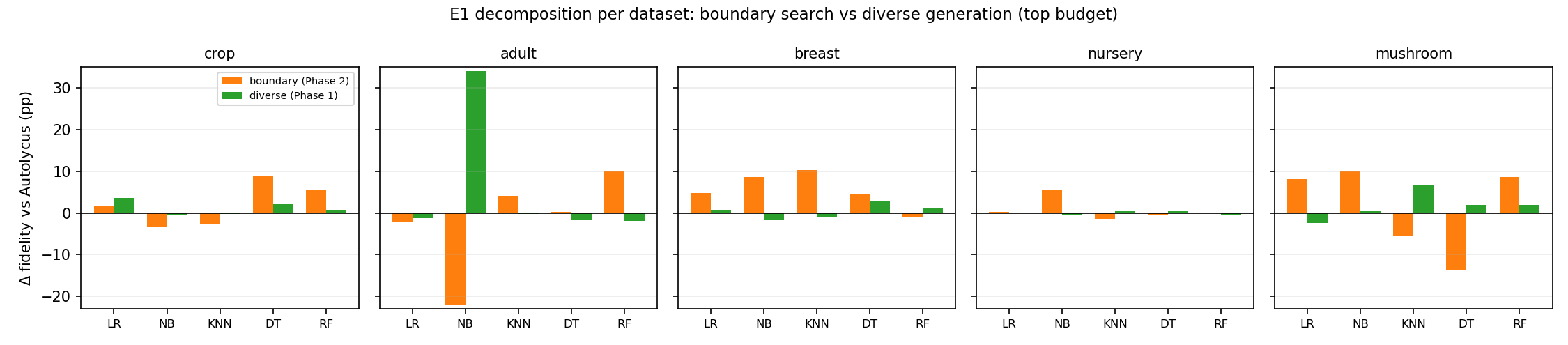}
\caption{Per dataset: contribution of boundary search (Phase~2, orange) and
diverse generation (Phase~1, green) to the LIME fidelity gain over Autolycus (percentage points, top
budget, mean over $S{=}10$ seed sets), one panel per dataset, grouped by target model. Shared
$y$-axis. The \textsc{adult}/NB diverse bar ($+34$pp) sets the scale.}
\label{fig:decomp-perdataset}
\end{figure*}

The split is also strongly dataset-dependent (Figure~\ref{fig:decomp-perdataset}): diverse
generation carries almost the entire gain on \textsc{adult}/NB ($+34$pp, a generative target starved
of coverage), while boundary search dominates on the tree targets of \textsc{crop} and the linear
targets of \textsc{breast}. Neither component is tied to a particular dataset, both being query placement mechanisms, which is why the aggregate in Figure~\ref{fig:decomp} splits evenly.

\section{Attack performance under matched information access}
\label{app:matched}
\label{subsec:e4}
This experiment runs our three phases with the target's explanation rather than a self-computed
grid, so both sides read the same \textsc{Lime} output and the comparison isolates query
placement. \S\ref{subsec:ladder} evaluates the
\sys\ attack, which is the one the paper proposes. This appendix and Appendix~\ref{app:decomp}
predate the multi-split protocol of \S\ref{subsec:setup} and are reported unchanged on the original
single split over five datasets, omitting pendigits, so that they stay comparable to the published
baseline.

\begin{figure*}[t]
\centering
\includegraphics[width=\textwidth]{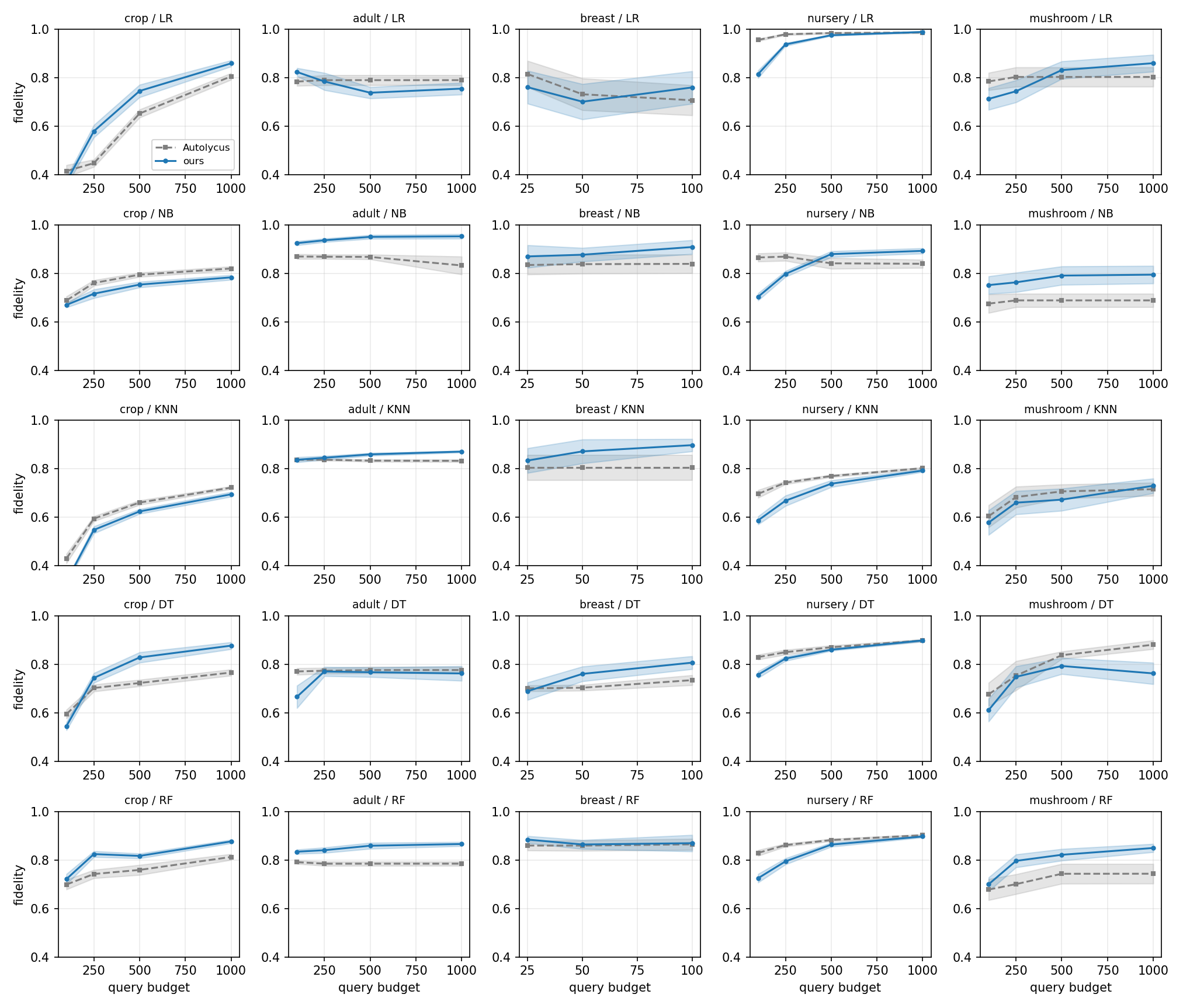}
\caption{Fidelity vs.\ query budget, our phases reading the target's explanation (solid, blue) vs.\ Autolycus-\textsc{Lime}
(dashed, grey), per target (rows) and dataset (columns). Band is the standard error over $S{=}10$ seed
sets. $n{=}1$, $k{=}3$, diverse budget fixed at cap${=}15$. Shared $y$-axis ($0.4$--$1.0$).}
\label{fig:lime-headline}
\end{figure*}

Figure~\ref{fig:lime-headline} plots surrogate fidelity against query budget for the
explanation-reading configuration and the Autolycus \textsc{Lime} baseline. At a single fixed diverse
budget it improves fidelity on 18 of 25 target/dataset combinations, with the most consistent gains
on the non-tree targets that have headroom (LR/NB on crop, nursery, and mushroom).

\paragraph{The same attack helps under SHAP}
Repeating this with \textsc{Shap} as the explainer ($n{=}5$, $k{=}3$) gives the same picture: the
same configuration improves over Autolycus-\textsc{Shap} on the same non-tree targets (LR/NB/KNN,
largest where the baseline has headroom, e.g.\ nursery and mushroom), while tree targets sit near
their fidelity ceiling and show no gain (Figure~\ref{fig:shap-headline}, Appendix~\ref{app:sweeps}).
That the same boundary-and-diversity machinery helps under both explainers, independent of the particular attribution scores it consumes, is the first indication that the attribution channel is
not what drives the gain (H1). The \textsc{Lime}-specific advantage of a free threshold is
isolated separately in \S\ref{subsec:e3}.

\paragraph{Diverse-budget sensitivity}
The effect of diverse generation depends on the target. It helps high-headroom targets and can hurt
near-ceiling or small continuous targets. We selected cap${=}15$ once in a global exploratory sweep
and fixed it for all reported target families and datasets. We did not tune this value per target or
target instantiation.

\section{Step-size sensitivity}
\label{app:steps}
The main \sys\ configuration uses the feature-scaled perturbation step $\varepsilon_i$, while the
published LIME traversal uses a unit step. We repeat the comparison with both methods using each
rule on the $20$ continuous-feature dataset--model cells. Within each method, switching the rule is
the only change. All configurations use the same $10$ train/test splits, retrained targets, seed
sets, attack random states, query budgets, evaluation sets, surrogate families, and refit counts.

For each cell, we first compute the mean paired fidelity difference over its $10$ target
instantiations. Table~\ref{tab:steps} averages these $20$ cell values. Its intervals are percentile
bootstrap intervals from $100{,}000$ resamples of the cells, so they summarize variation across the
evaluated cells rather than independent-dataset inference. In both direct, matched comparisons,
\sys\ has higher mean fidelity. Even against the higher-scoring Autolycus step rule selected within
each cell, the mean difference remains positive.

\begin{table}[t]
\centering
\caption{Step-size sensitivity on continuous-feature targets. Mean fidelity difference (pp) over
$20$ dataset--model cells.}
\label{tab:steps}
\begin{tabular}{lrr}
\toprule
contrast & mean & $95\%$ interval \\
\midrule
\sys\ $-$ Autolycus, feature-scaled & $+5.48$ & $[+1.55,+10.68]$ \\
\sys\ $-$ Autolycus, unit & $+3.96$ & $[+2.04,+5.86]$ \\
\sys\ $-$ stronger Autolycus rule & $+3.29$ & $[+0.94,+5.81]$ \\
Autolycus: feature-scaled $-$ unit & $+1.45$ & $[-3.77,+4.91]$ \\
\bottomrule
\end{tabular}
\end{table}

\section{How little auxiliary data rebuilds the grid}
\label{app:auxsize}
\S\ref{subsec:e5} shows that a discretizer the adversary fits on its own auxiliary partition
reproduces the target's. That partition is the ``comprehensive'' end of the range the threat model
inherits~\cite{oksuz2024autolycus}. Here we find the other end, sweeping the number of samples
per class the discretizer is fitted on while holding the attack, the seed set, the budget and
the RNG fixed, so that only the estimate of the grid varies. Table~\ref{tab:auxsize} gives the full
quartile-grid sweep. We run this on
the datasets with continuous features: on the fully categorical ones the bin edges are artefacts of
the integer encoding (\S\ref{sec:discussion}), so how fast that artefact converges is not informative.

\begin{table}[t]
\centering
\setlength{\tabcolsep}{4pt}
\caption{Full quartile-grid sensitivity sweep. Fidelity gain over no-bin-edge traversal (pp) with the
auxiliary grid fitted on $n$ inputs per class. The target column is the target-grid gain. Ten paired
splits.}
\label{tab:auxsize}
\begin{tabular}{llrrrrrr}
\toprule
 & & & \multicolumn{4}{c}{$\text{aux}$, fitted on $n$ per class} & \\
\cmidrule(lr){4-7}
dataset & model & target & $n{=}1$ & $n{=}2$ & $n{=}3$ & $n{=}5$ & full aux \\
\midrule
crop      & LR  & $+40.9$ & $+40.7$ & $+43.9$ & $+40.9$ & $+42.8$ & $+41.9$ \\
crop      & DT  & $+38.0$ & $+38.2$ & $+38.2$ & $+38.6$ & $+37.3$ & $+36.8$ \\
crop      & NB  & $+24.9$ & $+23.4$ & $+23.4$ & $+24.5$ & $+24.1$ & $+24.1$ \\
crop      & RF  & $+22.7$ & $+20.5$ & $+20.7$ & $+23.4$ & $+22.1$ & $+22.4$ \\
pendigits & LR  & $+18.6$ & $+12.9$ & $+15.4$ & $+16.5$ & $+16.6$ & $+20.1$ \\
crop      & KNN & $+15.2$ & $+10.9$ & $+16.4$ & $+17.1$ & $+15.9$ & $+17.7$ \\
breast    & NB  & $+11.2$ & $+6.0$  & $+10.9$ & $+9.7$  & $+11.9$ & $+11.9$ \\
pendigits & DT  & $+10.1$ & $+9.3$  & $+9.3$  & $+10.9$ & $+10.6$ & $+13.5$ \\
pendigits & NB  & $+7.9$  & $+5.2$  & $+7.8$  & $+7.8$  & $+6.8$  & $+7.8$ \\
\bottomrule
\end{tabular}
\end{table}

The requirement is very small, and crop and pendigits bound it from either side. On crop the adversary
is at parity with one sample per class: $+40.7$ against the target's $+40.9$ on LR, $+38.2$
against $+38.0$ on DT. Only KNN lags at $n{=}1$, and it catches up by $n{=}2$. Pendigits converges
more slowly, as its shape predicts: sixteen wide-range continuous features to place quartiles on and
ten classes rather than seventeen. Its tree and generative targets are nonetheless at parity almost at
once ($+9.3$ against the target's $+10.1$ on DT at one sample per class, $+7.8$ against $+7.9$ on NB
at two), while LR climbs $+12.9 \to +15.4 \to +16.5 \to +16.6$ and is still some two points short of
the target's $+18.6$ at five per class, closing only with the full partition. Breast sits between
them, at parity by $n{=}2$.

A quartile is a cheap statistic, and a handful of points per class estimates it well enough to recover
most of the gain: on crop the columns to the left of $\text{aux}$ are, to within noise, the same
numbers, and on pendigits they reach $83$ to $100$ per cent of it. An adversary holding a small
fraction of the auxiliary pool is therefore no worse off. Re-estimating the bins online from the
accumulated query set is the one variant that does worse, because the attack concentrates queries
near decision boundaries and its own history is therefore a biased estimator of the marginal
distribution the bins summarise.

\section{Class-support restriction sensitivity}
\label{app:auxshift}
The main RQ3 comparison varies the grid source while retaining the class-balanced seed protocol.
Here, the attacker instead receives data for only a random subset of classes. For crop, the moderate
and severe pools retain $11$ and $6$ of $17$ classes. For pendigits, they retain $7$ and $4$ of $10$.
The target and evaluation set retain all classes. In the direct source comparison, both configurations
start from the same restricted seed pool. One fits its grid on that pool and the other uses a grid
fitted on the target-training features. Thus, the target advantage in Table~\ref{tab:auxshift-cells}
does not include the loss from missing seeds.

\begin{table}[t]
\centering
\caption{Class-support sensitivity by target. Target-grid advantage (pp); positive values favour the target grid.}
\label{tab:auxshift-cells}
\setlength{\tabcolsep}{5pt}
\begin{tabular}{llrr}
\toprule
dataset & model & moderate & severe \\
\midrule
crop      & LR  & $-2.5$ & $+3.5$  \\
crop      & NB  & $+3.3$ & $+2.8$  \\
crop      & KNN & $+1.1$ & $+3.5$  \\
crop      & DT  & $-0.9$ & $+6.2$  \\
crop      & RF  & $+4.1$ & $+4.1$  \\
pendigits & LR  & $+2.8$ & $+6.5$  \\
pendigits & NB  & $+0.3$ & $+6.3$  \\
pendigits & DT  & $+1.2$ & $+10.5$ \\
\bottomrule
\end{tabular}
\end{table}

The moderate restriction has a small mean target advantage of $+1.18$pp, with six of eight cells
positive. Under the severe restriction, all eight cells favour the target grid and the mean advantage
is $+5.41$pp. These cells share datasets and splits, so their directions summarize the evaluated
settings rather than an independent-dataset significance test.

To separate this source effect from the cost of missing seeds, we form a $2\times2$ decomposition
using the full or restricted seed pool and the full or restricted attacker-side grid. Table~\ref{tab:auxshift-decomp}
reports fidelity differences over the full-seed, full-grid configuration. Restricting the seeds is the
dominant cost. Shifting the grid alone is small, but it compounds the seed restriction under the
severe condition. The restricted attacker still receives a large bin-edge gain over its matched
no-bin-edge traversal.

\begin{table}[t]
\centering
\caption{Class-support decomposition. Mean fidelity difference (pp) over eight cells.}
\label{tab:auxshift-decomp}
\setlength{\tabcolsep}{5pt}
\begin{tabular}{lrr}
\toprule
effect & moderate & severe \\
\midrule
restricted grid only  & $+0.1$  & $-1.3$  \\
restricted seeds only & $-9.3$  & $-20.2$ \\
restricted grid and seeds & $-11.0$ & $-26.7$ \\
interaction            & $-1.8$  & $-5.3$  \\
own-grid bin-edge gain & $+22.3$ & $+20.4$ \\
\bottomrule
\end{tabular}
\end{table}

\section{Query-budget sweeps}
\label{app:sweeps}
\begin{figure*}[tbp]
\centering
\includegraphics[width=\textwidth]{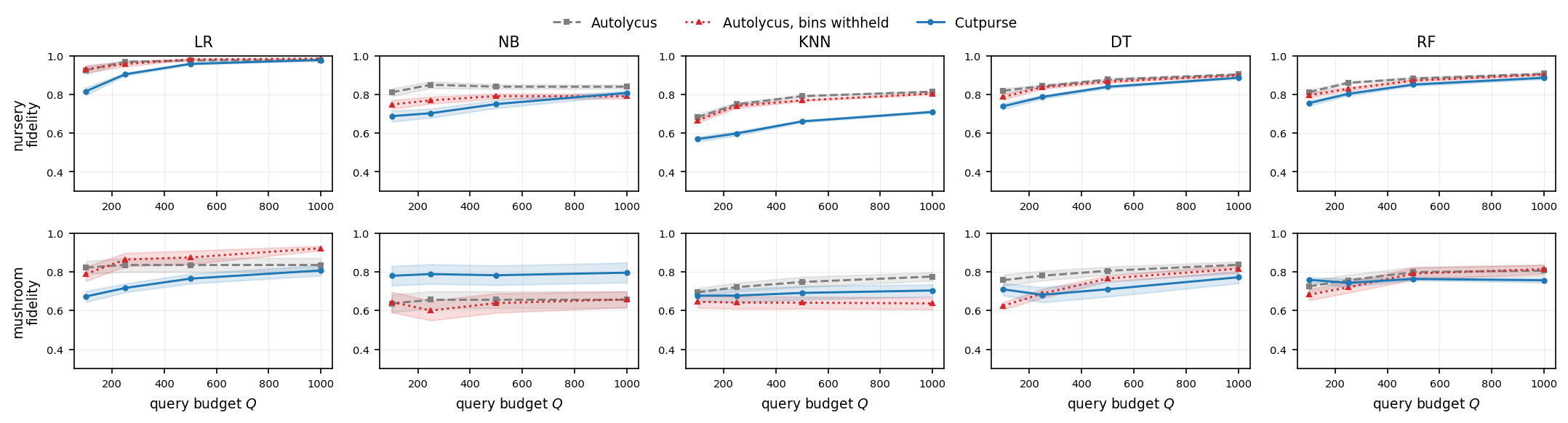}
\caption{Defense sweep on the two purely categorical datasets, the companion to
Figure~\ref{fig:defense-sweep}. Withholding the bin edges costs Autolycus almost nothing here, since
integer-encoded categories give the discretizer no meaningful cut points, and \sys\ has no edge to
snap to either.}
\label{fig:defense-sweep-categ}
\end{figure*}

This appendix collects two kinds of budget sweep. Figure~\ref{fig:defense-sweep-categ} is the
categorical companion to Figure~\ref{fig:defense-sweep} and, like it, evaluates \sys\ against the
defended and undefended service under the multi-split protocol. The remaining figures instead run our
three phases with the target's explanation rather than a self-computed grid, so that both sides read
the same output and the comparison isolates query placement. The \sys\ attack itself is
evaluated in \S\ref{subsec:ladder}.

\begin{figure*}[t]
\centering
\includegraphics[width=\textwidth]{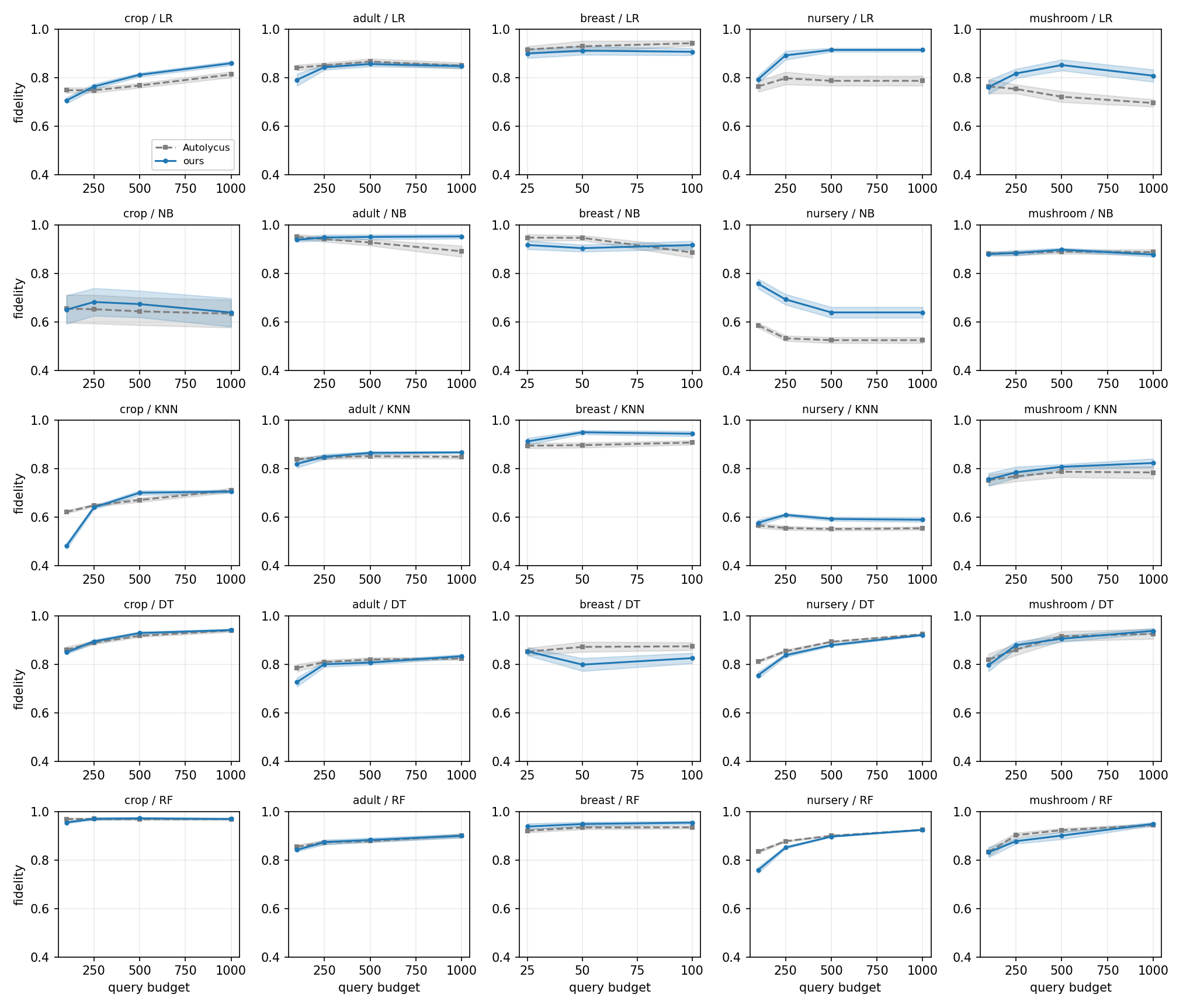}
\caption{\textsc{Shap} companion: surrogate fidelity vs.\ query budget. Our phases reading the target's \textsc{Shap} output
(solid, blue) vs.\ Autolycus-\textsc{Shap} (dashed, grey), per target (rows) and dataset (columns).
Shaded band is the standard error over the $S{=}10$ seed sets. $n{=}5$, $k{=}3$. All panels share a
fixed $y$-axis ($0.4$--$1.0$) for cross-panel comparability.}
\label{fig:shap-headline}
\end{figure*}

Figure~\ref{fig:shap-headline} is the \textsc{Shap} companion to Figure~\ref{fig:lime-headline},
discussed in \S\ref{subsec:e4}.

\end{document}